\UseRawInputEncoding

\documentclass[final,1p,times,onecolumn]{article}
\pdfoutput=1

\usepackage[numbers]{natbib}

\usepackage[utf8]{inputenc}

\usepackage{lineno}
\usepackage{float}
\usepackage{amssymb}

\usepackage{multirow}
\usepackage{adjustbox}
\usepackage{color}
\usepackage{amsmath}
\usepackage{dirtytalk}
\usepackage{mathtools}
\usepackage{textcomp}
\usepackage{verbatim}

\usepackage{hyperref}
\usepackage{silence}
\usepackage{tikz}
\usetikzlibrary{arrows.meta,calc,patterns,angles,quotes}
\usepackage{standalone}
\usepackage{amsmath}

\usepackage{geometry}

\newbox{\myorcidaffilbox}
\sbox{\myorcidaffilbox}{\large\includegraphics[height=1.7ex]{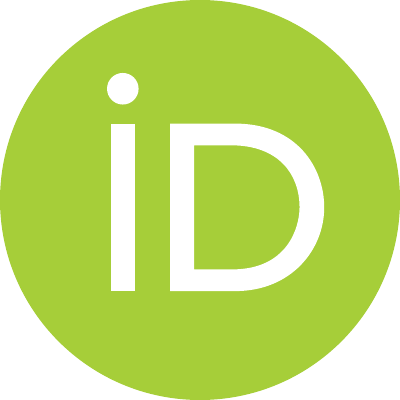}}
\newcommand{\orcidaffil}[1]{%
  \href{https://orcid.org/#1}{\usebox{\myorcidaffilbox}}}
  
\usepackage{caption}
\usepackage[figuresleft]{rotating}
\DeclareCaptionLabelFormat{continued}{#1~#2 (Continued)}

\usepackage[normalem]{ulem}

\newcommand{\rom}[1]{%
  \textup{\uppercase\expandafter{\romannumeral#1}}%
}

\usepackage[figuresleft]{rotating}
\usepackage{caption}

\usepackage{todonotes}

\usepackage{amssymb}

\usepackage{numcompress}

\newcommand{\graphicalabstract}[1]{
    \begin{figure}[h]
        \centering
        \includegraphics[width=\textwidth]{#1}
        \caption*{Graphical Abstract}
    \end{figure}
}

\title{Spheroidal harmonics for generalizing the morphological decomposition of closed parametric surfaces\\~\\\small{[Preprint]}}

\begin{document}

\author{Mahmoud Shaqfa \thanks{Massachusetts Institute of Technology (MIT), Department of Mechanical Engineering, Cambridge, MA 02139, USA--mshaqfa@mit.edu} \orcidaffil{0000-0002-0136-2391}
\and
Wim M. van Rees \thanks{wvanrees@mit.edu} \orcidaffil{0000-0001-6485-4804} \\
}

\date{}
\maketitle

\graphicalabstract{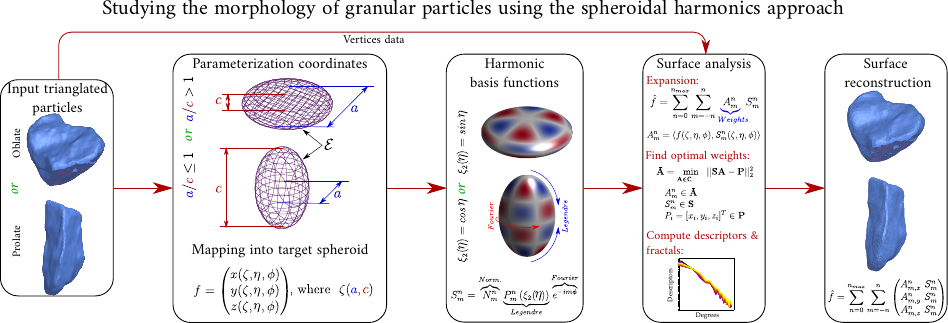}

\begin{abstract}
Spherical harmonics (SH) have been extensively used as a basis for analyzing the morphology of particles in granular mechanics. The use of SH is facilitated by mapping the particle coordinates onto a unit sphere, in practice often a straightforward rescaling of the radial coordinate. However, when applied to oblate- or prolate-shaped particles the SH analysis quality degenerates with significant oscillations appearing after the reconstruction. Here, we propose a spheroidal harmonics (SOH) approach for the expansion and reconstruction of prolate and oblate particles. This generalizes the SH approach by providing additional parameters that can be adjusted per particle to minimize geometric distortion, thus increasing the analysis quality. We propose three mapping techniques for handling both star-shaped and non-star-shaped particles onto spheroidal domains. The results demonstrate the ability of the SOH to overcome the shortcomings of SH without requiring computationally expensive solutions or drastic changes to existing codes and processing pipelines.
\end{abstract}


\section{Introduction}
\label{sec:intro}
\par
Accurate and efficient shape characterization of granular surfaces is crucial in many engineering applications, e.g., interparticle interaction \citep{Wang2023}, heterogeneous microstructures \citep{SHAQFA2022}, and fluid flow in granular media, to name just a few. These characterization processes yield shape signatures that can be used for classifying and comparing particulate matter \citep{Zhao2017}, as well as generating virtual microstructures with statistically accurate shapes.
\par
A frequently used approach for shape characterization relies on a harmonic decomposition of surfaces via the spherical harmonics (SH) analysis. This method is commonly used in biomedical applications and medical imaging, e.g., studying the shapes of the human brain \citep{Brechbuhler_1995}, heart ventricles \citep{Huang_2006}, and skulls \citep{giri2021open}. SH is also used in computer graphics and geometry processing \citep{Kazhdan2004} for comparing and classifying objects.
\par
\begin{figure}
    \centering
    \includegraphics[width=0.7\linewidth]{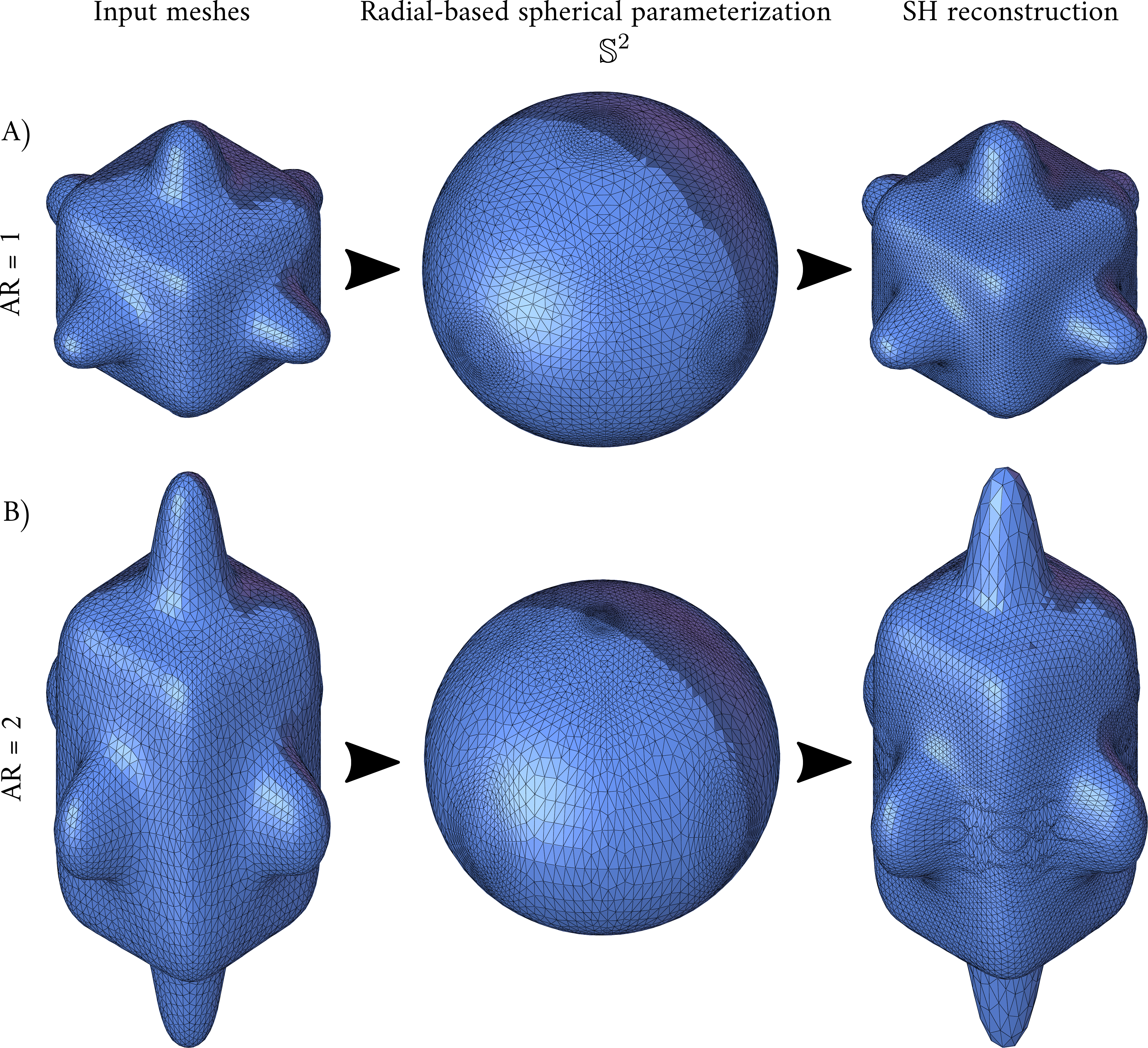}
    \caption{A benchmark shape [see the online documentation of \citep{libigl}] analyzed and reconstructed using the radial-based spherical harmonics approach. The top row (A) shows the original benchmark surface with an aspect ratio $\text{AR} = 1.0$, whereas the bottom row (B) repeats the analysis for the same surface stretched to an aspect ratio $\text{AR} = 2$. Each row shows, from left to right, the input meshes, the radial parameterization results, and the reconstruction results. The example shows how changing the AR away from unity distorts the mesh after the mapping, which limits the maximum reconstruction frequency and affects the orthogonality of the bases resulting in oscillations after reconstruction (bottom right).}
    \label{fig:intro_SH_rec}
\end{figure}

In the specific context of particulate and granular mechanics, SH analyses are used to characterize the shape of sand, gravel, and stones as well as to analyze and generate realistic virtual microstructures \citep{Qian_2012_thesis, Xiong2021, Wang2021, SHAQFA2022, Zheng2022, Huang2023, Zhao2023, Hu2024}. The surfaces in this domain are mostly (but not exclusively) star-shaped, so that an invertible mapping of surface coordinates to the sphere is typically obtained by simply normalizing the radial distance measured from the centroid of the particle surface (this approach is herein referred to as radial mapping). However, applying the SH approach to oblate- and prolate-shaped particles (e.g., river gravel and coral sand, respectively) reveals a visual degradation of the reconstruction quality using SH analysis. Figure \ref{fig:intro_SH_rec}, shows a comparison between reconstructing two artificial surfaces: one with an aspect ratio $\text{AR} = 1$ and the same surface but with $\text{AR} = 2$. The reconstruction of the second benchmark shows clear oscillations that take place in regions of large area distortions on the spherical image $\mathbb{S}^2$. Some examples of such oscillations in more practical scenarios are also visible in literature, such as \cite{Qian_2012_thesis} (Fig.~4.10 and 4.12 of that work) and \cite{Paixo_2021}. The degradation appears as high-frequency oscillations in localized regions of the reconstructed surfaces, and are thus difficult to capture with standard second norm errors. Nevertheless, the oscillations have an impact on the spectral properties that in turn affect the measure of the fractal dimension of the particles. Further, in mechanical simulations such as finite or discrete elements, these oscillations can impact the surface contact and the crack propagation along such interface \citep{SHAQFA2022}.
\par
Though the radial mapping approach for studying particles is prevalent due to its simplicity and low cost, other techniques for mapping surfaces onto spherical surfaces exist \citep{Choi2016, choi2020parallelizable}. However, these approaches do not completely mitigate the reconstruction problems for prolate/oblate geometries and the iterative nature of these algorithms comes with a high computational cost. For instance, using the Ricci flow approach \citep{Jin2008} with strict stopping criteria to conformally map a surface onto a sphere can result in large area distortion that might lead to large gaps in the distribution of the points. These gaps are limited with bandwidth, so the higher harmonics (frequencies) do not converge at the affected regions \citep{McEwen2011} which will be a source of oscillations at that location. Recently, \cite{Huang2023} attempted to address the oscillation problem by proposing a \say{shrinkage strategy} to limit the effect of amplified zeroth-order harmonics. This procedure resulted in fewer oscillations in the middle band of the particles, but this approach depends on both a rescaling of the stones and numerical damping of the harmonic basis and thus affects the shape descriptors. %
\par
The appearance of these spurious oscillations on surfaces of low true sphericity index \citep{Wadell1932} implies that the prevalent mapping approaches to obtain spherical coordinates induce large topological angle and/or area distortions. This distortion causes discrepancies in the distribution of points on the sphere. Consequently, the orthogonality of the SH basis is increasingly lost. The errors in the reconstruction get worse as the number of analysis degrees increases, demonstrating that this is a divergence issue where Fourier weights do not tend to diminish for high degrees. Besides the mapping approach, other aspects can play a role in the reconstruction errors for complex geometries. The regularization step is used to estimate Fourier coefficients of discrete noisy data parameterized onto a specific domain, and some regularization methods are known sources of reconstruction errors. However, these errors are well understood and can be mitigated by using a proper scalable algebraic orthogonal projection method \citep{Shaqfa2023OnMethod}. Further, Gibbs oscillations may appear for geometries with sharp edges \citep{Gelb_1997}. Combined, mapping-induced distortions are a dominant source of reconstruction errors for non-spherical particles, for which simple successful solution strategies have not yet been formulated. 
\par
A possible avenue to mitigate the mapping-induced distortions is to define a parameterization domain that more closely adheres to the dominant geometric features of the considered shapes. Along this line, recent generalizations of the original SH method have taken place to account for open surfaces using the hemispherical harmonics (HSHA) approach \cite{Huang_2006, Giri_2021}. This was followed by a more general approach of open surfaces through the spherical cap harmonics (SCHA) \citep{shaqfa2021b}. The underpinning idea in the SCHA method is to analyze open surfaces on more general and customizable parameterization domains (i.e., spherical caps prescribed by an opening angle $\theta_c$). More recently, we also proposed a more computationally efficient approach based on the disk harmonics analysis (DHA) to treat nominally flat self-affine rough surfaces \citep{SHAQFA2023b}. Using such custom parameterization domains in SCHA and DHA to lower mapping-induced topological distortions improves the decomposition and reconstruction results.
\par
Motivated by the success of customizable mapping domains, we propose here to use confocal spheroidal coordinates (oblate and prolate spaces) for the characterization of oblate/prolate granular particles. These domains are customizable in that they contain parameters (the major and minor axes of the spheroid) to be tuned for each particle separately. Formally, they generalize the SH coordinates and contain the SH as a special case. In fact, similar to the SH, the spheroidal basis functions consist of the associated Legendre polynomials with Fourier bases, which makes the transition to spheroidal harmonics seamless for the granular mechanics community. Further, previous experimental work uses spheroids to classify and characterize a wide spectrum of granular particles \citep{Zhou2011, Zhu2017}. This implies that even with a simple low-cost mapping technique such as the radial mapping commonly used in the granular mechanics community, one can expect reduced distortions in this domain.
\par
The rest of this paper is organized as follows. Section \ref{sec:spheroidal_harm} introduces the spheroidal coordinate system and the harmonic bases used for the analysis. Section \ref{sec:param} is where we propose methods to determine the spheroidal domains and proper parameterization methods for the analysis of star-shaped (SS) and non-star-shaped (NSS) closed surfaces. Detailed results and comparisons are depicted in Section \ref{sec:results}. Eventually, conclusions and future works are summarized in Section \ref{sec:conclusions}. For reproducibility, the Python3 codes that were used in this paper are made available under an open-source license.

\section{Spheroidal space and the corresponding harmonics}
\label{sec:spheroidal_harm}

\subsection{Ellipsoidal and spheroidal coordinates}

\par
\begin{figure}
    \centering
    \includegraphics[width=1.0\linewidth]{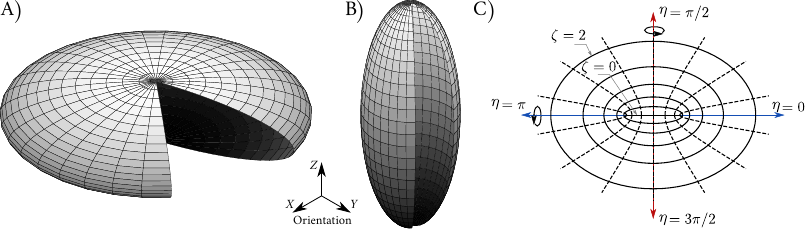}
    \caption{Oblate (A) and prolate (B) spheroids $\mathcal{E}$ embedded in $\mathbb{R}^3$. (C) The elliptic coordinates where the solid lines represent the confocal ellipses while the dashed lines are the confocal hyperbolas. When the elliptic system revolves about the vertical axis (red) it generates an oblate surface, while the prolate one can be generated by revolving about the horizontal axis (blue). %
    }
    \label{fig:oblate_vs_prolate}
\end{figure}

An ellipsoidal surface in $\mathbb{R}^3$ and centered about the origin $\mathcal{O}$ is written in the Cartesian coordinates as:
\begin{equation}\label{eqn:ellipsoid}
    \frac{x^2}{a^2} + \frac{y^2}{b^2} + \frac{z^2}{c^2} = 1,
\end{equation}
where $a, b, c$ are positive real numbers that describe the semi-axis lengths along each of the coordinate directions. Spheroidal surfaces form a subset of \eqref{eqn:ellipsoid} where $a = b$, and can thus be written as:
\begin{equation} \label{eqn:spheroid}
    \mathcal{E}: \frac{x^2 + y^2}{a^2} + \frac{z^2}{c^2} = 1,
\end{equation}
More generally, in this work, we always interpret $a$ as the repeated semi-axis length and $c$ as the third (non-repeating one), independent of the orientation of the spheroid in $\mathbb{R}^3$. The object is classified as oblate if $a > c$ (Fig.~\ref{fig:oblate_vs_prolate}A) or as prolate if $c \geq a$ (Fig.~\ref{fig:oblate_vs_prolate}B). In the latter condition, we intentionally included the sphere ($a = b = c$) because the prolate harmonics mathematically degenerate to the spherical ones as $c \to a$ (as discussed below). Further, we denote the aspect ratio $\text{AR} = a/c$ (i.e., $a/c > 1$ is for oblate spheroids, $a/c < 1$ is for prolate ones, and $a/c = 1$ is for a sphere).

Spheroidal coordinates are obtained by revolving elliptic coordinates around an axis. The elliptical coordinates (shown in Fig.~\ref{fig:oblate_vs_prolate}C) are an orthogonal system formed by confocal ellipses (solid lines) and hyperbolae (dashed lines). Here any elliptical section is identified by $\zeta$ $\in \mathbb{R}^+$, which is analogous to the radial vector in polar coordinates, and the hyperbolic section is identified by the latitude angle $\eta$, analogous to the angular coordinate in polar coordinates. 
\par
In Fig.~\ref{fig:oblate_vs_prolate}C, revolving the depicted elliptic coordinates about the horizontal axis generates prolate coordinates, while revolving it about the vertical axis generates oblate ones. The azimuthal angle $\phi$ $\in [0,2\pi)$ then completes the spheroidal coordinate system $(\zeta, \eta, \phi)$. Spheroidal coordinates form quadric surfaces of revolution that define a complete set of coordinates to sufficiently describe any point in $\mathbb{R}^3$.

\par
Starting with oblate spheroids, the parametric form of the oblate spheroidal coordinates can be written as:
\begin{equation}\label{eqn:oblate_para1}
    \begin{aligned}
        x(\zeta, \eta, \phi) & = e \ \cosh \zeta \ \cos \eta \ \cos \phi,
        \\ y(\zeta, \eta, \phi) & = e \ \cosh \zeta \ \cos \eta \ \sin \phi,
        \\ z(\zeta, \eta) & = e \ \sinh \zeta \ \sin \eta,
    \end{aligned}
\end{equation}
where $(\pm e, 0)$ are the coordinates of the foci along the major axis, and the latitude angle $\eta \in [-\pi/2, \pi/2]$. Similarly, the parametric representation of the prolate coordinates can be written as:
\begin{equation}\label{eqn:prolate_para1}
    \begin{aligned}
        x(\zeta, \eta, \phi) & = e \ \sinh \zeta \ \sin \eta \ \cos \phi,
        \\ y(\zeta, \eta, \phi) & = e \ \sinh \zeta \ \sin \eta \ \sin \phi,
        \\ z(\zeta, \eta) & = e \ \cosh \zeta \ \cos \eta.
    \end{aligned}
\end{equation}
For the prolate coordinates the latitude angle $\eta \in [0, \pi]$. 
\par
In the context of this paper, we focus primarily on spheroidal surfaces, which are defined by a constant value of $\zeta = \zeta_0$ and parametrized solely through $\eta$ and $\phi$. From the parametric forms in Eq.~\eqref{eqn:oblate_para1} and \eqref{eqn:prolate_para1} we can write the aspect ratio AR of any spheroidal surface as a function of $\zeta_0$, where $\text{AR} = [\tanh \zeta_0]^{-1}$ for oblates or $\text{AR} = \tanh \zeta_0$ for prolates.

\subsection{Spheroidal harmonic functions}
\par
The harmonics of oblate and prolate spheroidal coordinates are a special case of the ellipsoidal harmonics, which generalize the spherical harmonics. Similar to the spherical harmonics, the basis can be obtained by assuming that the differential form of the elliptic partial differential equation of Laplace $\nabla^2 f = 0$ is separable, where $\nabla^2 = \Delta$ is the Laplace-Beltrami operator. In this section, we briefly describe the derivation of the bases and introduce our notation to align it with the used SH notation in the granular mechanics literature. Further details can be found in standard textbooks, e.g., \cite{Byerly_1893, smythe1950, Morse_Feshbach_1953, abramowitz_stegun_1964, Moon1988}.

\subsubsection{Basis functions of oblate coordinates}
\label{sec:oblate_basis}
\par
The Laplacian expression can be simplified by letting $ \xi_1 = \sinh{\zeta} $ and $ \xi_2 = \sin{\eta} $, so that $\xi_1 \in [0, \infty)$ and $\xi_2 \in [-1, 1]$ \citep{smythe1950}. By using the identities $\cosh^2{a} - \sinh^2{a} = 1$ and $\cos^2{b} + \sin^2{b} = 1$, we can rewrite the parametric coordinates in Eq. \eqref{eqn:oblate_para1} as:
\begin{equation}
    \begin{aligned}
        x(\xi_1, \xi_2, \phi) & = e \ \sqrt{(1+\xi^2_1) (1-\xi^2_2)} \ \cos \phi,
        \\ y(\xi_1, \xi_2, \phi) & = e \ \sqrt{(1+\xi^2_1) (1-\xi^2_2)} \ \sin \phi,
        \\ z(\xi_1, \xi_2) & = e \ \xi_1 \ \xi_2.
    \end{aligned}
\end{equation}
\par
After writing the metric coefficients in the curvilinear coordinates, we can write the Laplacian operator in oblate coordinates as:
\begin{equation}\label{eqn:full_laplace_oblate}
    \nabla^{2} f = \frac{1}{e^{2}(\xi_1^{2}+\xi_2^{2})}\left(\frac{\partial}{\partial\xi_1}\left((\xi_1^{2}+1)\frac{\partial f}{\partial\xi_1}\right)+\frac{\partial}{\partial\xi_2}\left((1-\xi_2^{2})\frac{\partial f}{\partial\xi_2}\right)+\frac{\xi_1^{2}+\xi_2^{2}}{(\xi_1^{2}+1)(1-\xi_2^{2})}\frac{{\partial f}^{2}}{{\partial\phi}^{2}}\right).
\end{equation}
In our context, we only consider spheroidal surfaces where $\xi_1 = const$, so we drop all the terms related to $\xi_1$ from the general solution to obtain the Laplace-Beltrami operator on an oblate spheroidal surface.
\par
Classically, the solution of Eq.~\eqref{eqn:full_laplace_oblate} can be obtained by separation. The general solution form can be written as:
\begin{equation}
   f(\eta, \phi) = \ \Theta \Big( \xi_2 (\eta) \Big) \ \Phi(\phi).
\end{equation}
\par
Substituting the general solution into \eqref{eqn:full_laplace_oblate}, we obtain two equations. The first is Euler's ordinary differential equation (ODE) as a function of $\phi$ only:
\begin{equation} \label{eqn:sep_fourier_ode}
    \frac{1}{\Phi} \frac{{\partial^{2} \Phi}}{{\partial\phi}^{2}} = - m^2,
\end{equation}
with $m^2$ is any arbitrary positive number. The second ODE is obtained from applying the separation of variables twice on the general form of Eq.~\eqref{eqn:full_laplace_oblate} and bundling the rest of the terms as a function of $\xi_2$ only:
\begin{equation} \label{eqn:sep_oblate_ode2}
    \frac{\partial}{\partial\xi_2}\left((1-\xi_2^{2})\frac{\partial \Theta}{\partial\xi_2}\right) - \frac{\Theta \ m^2}{1-\xi_2^{2}} + \Theta \ n(n+1) = 0.
\end{equation}
Equation \eqref{eqn:sep_oblate_ode2} is Legendre's associated differential equation of order $m$ and degree $n$. Another form of Eq.~\eqref{eqn:sep_oblate_ode2} can be commonly found in the literature by applying the exterior derivative on the first term inside the parenthesis of the left-hand side.
\par
Now, we can write the particular solutions of both ODEs. The solution for Eq.~\eqref{eqn:sep_fourier_ode} can be expressed via Fourier series:
\begin{equation}
    \Phi_m(\phi) = e^{\pm i m \phi},
\end{equation}
where $i^2 = -1$. Since the solution along $\phi$ is periodic [with $\Phi_m(\phi) = \Phi_m(\phi+2\pi)$ and $\Phi_m^{\prime}(\phi) = \Phi_m^{\prime}(\phi+2\pi)$], $m$ must be an integer.
\par
The solution for Eq.~\eqref{eqn:sep_oblate_ode2} can be expressed with the first- and second-kind associated Legendre polynomials, $P_m^n (\xi_2)$ and $Q_m^n (\xi_2)$ respectively. However, here we only use the first kind as the second kind is singular at the poles of the oblate coordinates. From Eq.~\eqref{eqn:sep_oblate_ode2}, we write the solution as a function of $\eta$:
\begin{equation}
    \Theta_m^n = P_m^n(\sin{\eta}).
\end{equation}
\sloppy Now, the particular solution can take the form $f^{n}_{m} = P_m^n(\sin{\eta}) \ e^{i m \phi}$ for arbitrary $n$ and $m \in \{-n, \dots, 0, \dots, n\}$. The degrees $n \in \mathbb{Z}^+$ represent the ranked roots (eigenvalues) of the associated Legendre polynomials. Subsequently, we can expand a continuous field $f$ distributed uniformly over  an oblate as a linear series with weights $A_m^n$ and normalization factors $N_m^n$:
\begin{equation}\label{eqn:general_sol_oblate}
    f(\eta, \phi) = \sum_{n=0}^{\infty} \sum_{m=-n}^{n} N_m^n \ A_m^n \ P_m^n(\sin{\eta}) \ e^{i m \phi}. 
\end{equation}

\subsubsection{Basis functions of prolate coordinates}
\label{sec:prolate_basis}
\par
As above, let $\xi_1 = \cosh{\zeta}$ and $\xi_2 = \cos{\eta}$; with $\xi_1 \in [0, \infty)$ and $\xi_2 \in [-1, 1]$. Then, we can rewrite the prolate parametric coordinates in Eq.~\eqref{eqn:prolate_para1} as:
\begin{equation}
    \begin{aligned}
        x(\xi_1, \xi_2, \phi) & = e \ \sqrt{(\xi^2_1 - 1) (1-\xi^2_2)} \ \cos \phi,
        \\ y(\xi_1, \xi_2, \phi) & = e \ \sqrt{(\xi^2_1 - 1) (1-\xi^2_2)} \ \sin \phi,
        \\ z(\xi_1, \xi_2) & = e \ \xi_1 \ \xi_2.
    \end{aligned}
\end{equation}
\par
The corresponding general Laplacian operator of prolate coordinates can be written as:
\begin{equation} \label{eqn:full_laplace_prolate}
    \nabla^{2} f = \frac{1}{e^{2}(\xi_1^{2}-\xi_2^{2})}\left(\frac{\partial}{\partial\xi_1}\left((\xi_1^{2}-1)\frac{\partial f}{\partial\xi_1}\right)+\frac{\partial}{\partial\xi_2}\left((1-\xi_2^{2})\frac{\partial f}{\partial\xi_2}\right)+\frac{\xi_1^{2}-\xi_2^{2}}{(\xi_1^{2}-1)(1-\xi_2^{2})}\frac{{\partial^{2} f}}{{\partial\phi}^{2}}\right).
\end{equation}
From Eq.~\eqref{eqn:full_laplace_prolate} we can see that the general form is similar to Eq.~\eqref{eqn:full_laplace_oblate} apart from a change in sign in some terms and the definition of $\xi_1$ and $\xi_2$. This only changes the particular solution that depends on $\eta$, which now becomes:
\begin{equation}
    \Theta_m^n = P_m^n(\cos{\eta}).
\end{equation}
With this, we can write the expansion of any field $f$ as an infinite series:
\begin{equation}\label{eqn:general_sol_prolate}
    f(\eta, \phi) = \sum_{n=0}^{\infty} \sum_{m=-n}^{n} N_m^n \ A_m^n \ P_m^n(\cos{\eta}) \ e^{i m \phi},
\end{equation}
where $m$ is the order and $n$ is the degree of the solution. %

\subsubsection{Summary of the spheroidal harmonics}
\par
Here we summarize the spheroidal basis for any degree $n$ and order $-n\leq m \leq n$. The number of evaluated basis functions up to any $n$-degree is $(n+1)^2$. The normalization factors $N_m^n$ ensure orthonormality of the basis and match the ones derived for the spherical harmonics \citep{abramowitz_stegun_1964}:
\begin{equation}
    N_m^n = \sqrt{\frac{(2n+1)}{4\pi} \ \frac{(n-m)!}{(n+m)!}}.
\end{equation}
\par
\begin{figure}
    \centering
    \includegraphics[width=0.7\linewidth]{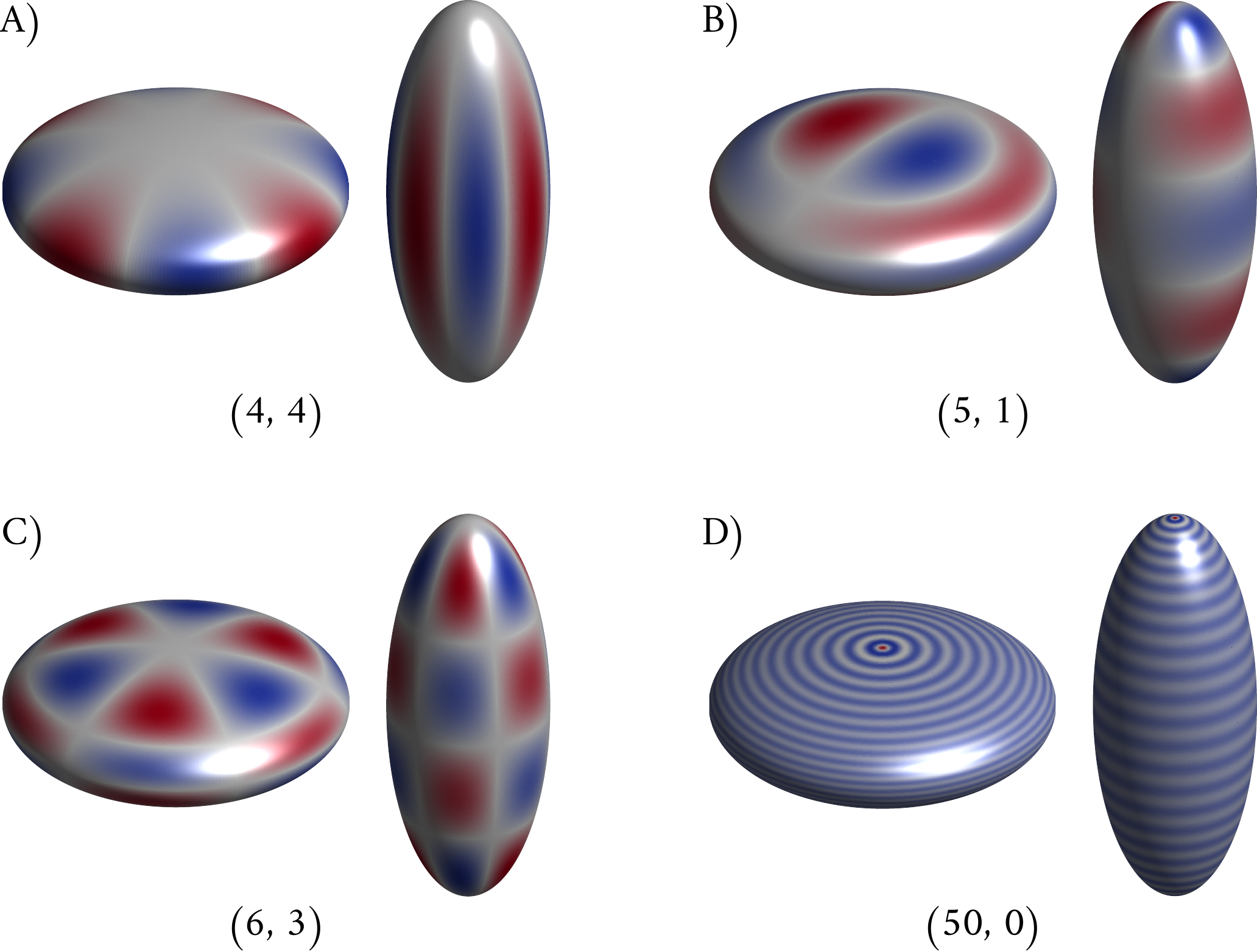}
    \caption{Examples of the oblate and prolate harmonic basis. (A) $\to$ (D) contain oblate (left) and prolate (right) harmonics that correspond for $(n, m)$ pairs (below). To construct the spheroids, we used $\zeta = 0.4$ and $e = 5$. The color represents the normalized real part of the basis $N_m^n \ \Re\{\Theta_m^n(\eta) \ \Phi_m(\phi)\}$.
    }
    \label{fig:rendering_basis}
\end{figure}

Any function $f$ defined on a spheroidal surface of constant $\zeta$ can then be decomposed and approximated as a truncated series up to $n_{max}$ degrees as:
\begin{equation}\label{eqn:general_sol}
\hat{f}(\eta, \phi) = \sum_{n=0}^{n_{max}} \sum_{m=-n}^{n} N_m^n \ A_m^n \ P_m^n\left(\xi_2(\eta)\right) \ e^{i m \phi}, \text{ where}
\end{equation}
\begin{equation}
\xi_2(\eta) = 
\begin{cases}
  \sin \eta & a/c > 1 \\
  \cos \eta & a/c \leq 1
\end{cases}.
\end{equation}
Notice that $\xi_2 (\eta)_{oblate} = \xi_2 (\eta + \pi/2)_{prolate}$. Numerically, this is equivalent to the spherical harmonics basis but wrapped over a different domain. Due to the truncation error with the infinite series, we used $\hat{f}$ in \eqref{eqn:general_sol} instead of $f$ such that $\hat{f} \to f$ as $n_{\max} \to \infty$. To efficiently solve for the Fourier weights $A_m^n$ we refer the readers to the newly proposed approach by \cite{Shaqfa2023OnMethod} through the modified randomized Kaczmarz (RK) algorithms that were accelerated with the conjugate symmetry and sparsity of the signals. Lastly, figure~\ref{fig:rendering_basis} visualizes the real part of the harmonics for oblate and prolate basis.

\section{Surface parameterization onto spheroidal domains}
\label{sec:param}
\par
The surface parameterization bijectively maps surfaces that are embedded in the Cartesian space $\mathbb{R}^3$ onto a target 2-manifold. These mappings are usually not isometric as they do not preserve distances locally. Thus, we can classify mappings into angle-preserving (conformal) and area-preserving mappings. Conformal mapping does not, in general, preserve the local areas of the mapped surfaces which might result in large area distortions. Conversely, area-preserving mapping does not preserve angles. The choice of a suitable mapping is generally application-dependent and keeping a balance between the area- and angle-preserving approaches is often desired \citep{Nadeem2017}.
\par
The surface parameterization step is essential to find an equivalent embedding of the surface into the analysis domain. The most common particle-related SH algorithms use radial mapping into a unit sphere, e.g., \cite{Qian_2012_thesis, Paixo_2021}. Although this approach is not guaranteed to be bijective and often induces both angle and area errors, it is simple and numerically efficient, which is important when hundreds of particles are being analyzed in granular mechanics applications.
\par
In this section, we first introduce two possible coordinate inversions from $\mathbb{R}^3$ onto $(\eta, \phi)$ that underlie two of the surface parameterizations used in this work. Both inversions are parameterized, and we subsequently discuss the choice of the domain parameters as well as a transformation strategy to register surfaces into a canonical coordinate system. These two parameterizations extend the traditional radial projection in SH, but can only be used for star-shaped (SS) particles (see \ref{app:bijective_mapping} for a definition). To handle also non-star-shaped (NSS) particles, this section finally discusses a preprocessing step based on the \textit{conformalized mean curvature flow} (cMCF) method \citep{Kazhdan_2012}. Using this method we can flow particles into NSS particles into SS ones and subsequently use either of the two parameterization techniques.

\subsection{Coordinate inversions}
\label{subsec:inverse_coords}
\par
The main purpose of this section is to compute the parameterization coordinates in the general solution of Eq.~\eqref{eqn:general_sol}; we need to find $\phi$- and $\eta$-coordinates that correspond to the surface $(x, y, z)$ coordinates. As spheroids are surfaces of revolution (see Fig.~\ref{fig:oblate_vs_prolate}), computing $\phi$ is unambiguous and straightforward, such that the inverted azimuthal coordinate $\phi = \tan^{-1} (y/x)$. This definition is similar to the one used for spherical coordinates. For the rest of this section, we only discuss methods to determine the $\eta$-coordinate.
\par
First, it is natural to use the curvilinear spheroidal coordinates to define the mapping onto the target spheroids, and we denote this approach the \textit{hyperbolic mapping}. In \ref{APP:inverse_coordinates} we describe the native analytic derivation of the inversion of the curvilinear spheroidal coordinates to compute $(\zeta, \eta, \phi)$ of a given surface points $\in \mathbb{R}^3$. The analytic coordinate inversions for $\eta$, for oblate and prolate spheroids respectively, are given as: 
\begin{equation}\label{eqn:inv_ob}
    \eta = \Im\left\{\cosh^{-1}\left( \frac{\rho_{ob} + i z}{e}\right)\right\}, \quad
    \zeta = \Re\left\{\cosh^{-1}\left( \frac{\rho_{ob} + i z}{e}\right)\right\}.
\end{equation}
\begin{equation}\label{eqn:inv_pr}
    \eta = \Im\left\{\cosh^{-1}\left( \frac{i \rho_{pr} + z}{e}\right)\right\},
    \quad
    \zeta = \Re\left\{\cosh^{-1}\left( \frac{i \rho_{pr} + z}{e}\right)\right\}.
\end{equation}
Where $\rho = \sqrt{x^2 + y^2}$ is computed for oblate or prolate spheroids as shown in subscripts of $\rho$ in Eq.~\eqref{eqn:inv_ob} and \eqref{eqn:inv_pr} respectively. In this inversion, the value of $\zeta$ will not affect the hyperbolic coordinate $\eta$; the choice of $e$ is the free domain parameter that can be adjusted to each particle individually, as discussed in the next section. This is analogous to spherical harmonics where the value of the radial coordinate $r$ is invariant to the basis functions for surface shape morphology. Further, note that as $e \to 0$, the values of $\eta$ become equivalent to the polar coordinates $\theta$, and the mapping will reduce to a traditional radial mapping as used in SH.
\par
\begin{figure}[!ht]
    \centering
    \includegraphics[width=0.99\linewidth]{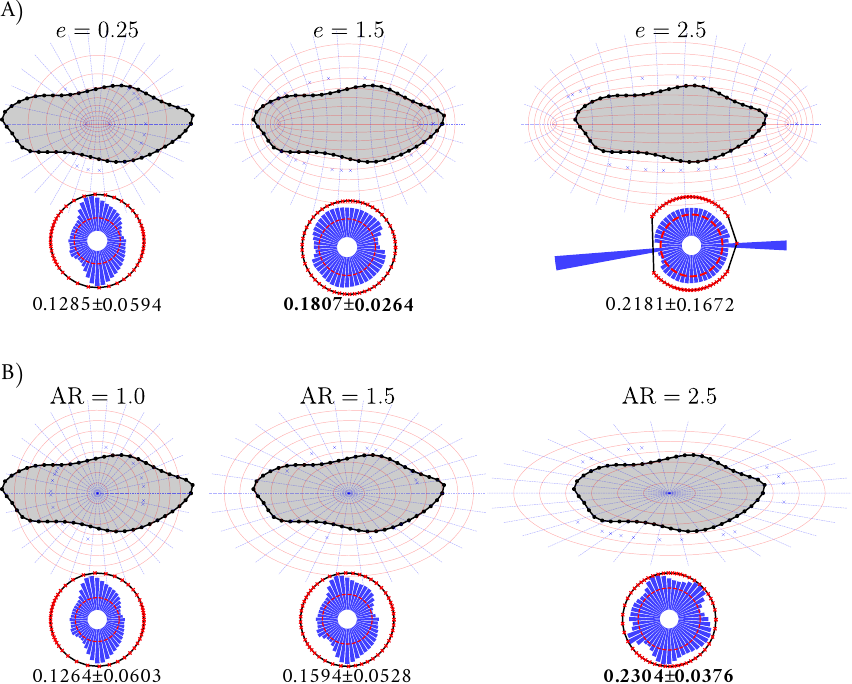}
    \caption{Computing the mapping of 2D closed contours, constituted by equidistant segments, via two different coordinate inversions. The 2D contours were used to simplify the visualization of the mapping process as a cross-section of a given stone. (A) mapping the 2D contour using hyperbolic mapping onto several target coordinates as a function of the focal distance $e$. For this set the choice of $\zeta$ was arbitrary as the mapping is a function of $e$ only. (B) mapping of the same contour using the rescaled polar coordinates approach as a function of the aspect ratio $\text{AR} = a/c$. The radial bar chart underneath each inset shows the radial distribution of the arclength between two consecutive points onto the target elliptic domain and was annotated with the average $\pm$ standard deviation (STD). The confocal red dashed circle in the radial charts marks the corresponding average arclength. Minimal STD (\textbf{boldfaced}) of the illustrated examples represents a more uniform mapping that typically corresponds to a better reconstruction accuracy. The mapped points (red crosses), in the lower insets, have been visualized when $\zeta \to \infty$ for the hyperbolic mapping (A), and rescaled circles with $\text{AR} \to 1$ for the radial one (B).} \label{fig:linear_vs_nonlinear_mapping_intro}
\end{figure}

Figure \ref{fig:linear_vs_nonlinear_mapping_intro}A shows the hyperbolic mapping in 2D for different choices of the domain parameter $e$. From the same figure, it can be seen how the distances between the mapped points (red crosses) onto the target ellipse change; thus, directly affecting the distribution of $\eta$. The sampling of $\eta$ is critical to the orthogonality of the basis functions and the overall reconstruction quality as entailed in Eq.~\eqref{eqn:general_sol}. In Fig.~\ref{fig:linear_vs_nonlinear_mapping_intro}A when $e = 2.5$ the focal points are laid outside the input contour, and as a result, the mapping results have large gaps in the $\eta$ coordinate (trimmed around the poles). These gaps will result in badly conditioned basis functions that degrade the reconstruction quality. In the next sections, we will describe how to choose the parameter $e$ to avoid this.
\par
Second, we consider a coordinate inversion based on rescaling the conventional spherical coordinates according to a certain chosen aspect ratio, which is the free parameter of this inversion. Rescaling the unit sphere with aspect ratio AR leads to a spheroid of constant $\zeta = \zeta_0$, with $\text{AR} = [\tanh \zeta_0]^{-1}$ for oblates or $\text{AR} = \tanh \zeta_0$ for prolates (where the scaling is independent of $e$). To compute the coordinate inversion from given particle surface points $\in \mathbb{R}^3$, we then use a standard spherical radial projection to compute an approximation to the latitude angle denoted as $\hat{\eta}$. This simplified radial mapping replaces the exact hyperbolic coordinate $\eta$ with the approximation: 
\begin{equation}\label{eqn:appro_inv}
    \hat{\eta} = \tan^{-1}\left(\text{AR} \, \frac{z}{\sqrt{x^2 + y^2}}\right).
\end{equation}
This approximation can be understood mathematically as a linearization (first-order approximation) of an exact spheroidal coordinate $\eta$, discarding higher-order terms in the distance from the target surface with $\zeta_0$ and $e$ (a full derivation is shown in \ref{app:linearized_inverse_coordinates}). As $\text{AR}\to 1$, this mapping is equivalent to the radial mapping used in SH.
\par
Figure \ref{fig:linear_vs_nonlinear_mapping_intro}B shows an example of this 2D radial mapping approach onto different elliptic targets for multiple choices of the domain parameter AR. Due to its similarity with the traditional radial mapping in SH, we here refer to it as the radial mapping approach.

\subsection{Choice of domain parameters}
\label{subsec:fitting_spheroid}
\par
Here we discuss the choice of the domain parameters $e$ (for the hyperbolic mapping) and $\text{AR}$ (for the radial mapping), starting from an arbitrary closed surface with $n_p$ vertices in $\mathbb{R}^3$.
First, we discuss a generic transformation to place any particle object in canonical coordinates. Then, we discuss the choice of domain parameters and the details of the mapping for each of the three proposed approaches.

\subsubsection{Canonical coordinates}
\par
To reduce the dimensionality of the fitting problem and place objects in canonical coordinates, we first translate our object such that the geometrical centroid is placed at the origin $\mathcal{O}$. Next, we rotate the object to align the object axis that maximizes the variance (largest dimension of the object) with the $X$-axis, while the one with a minimum variance is aligned with the $Z$-axis. One efficient way to do that is by applying the singular value decomposition (SVD) approach to the $3 \times n_p$ vertex matrix $P$, yielding $P = U \Sigma V^{T}$ with $U$ is a $3 \times 3$ rotation matrix. The canonical placement can then be computed as $\hat{P} = U^{T} P$. By construction, prolate-like surfaces will have their major axis aligned over the $X$-axis. So, we rotate prolates by an additional $+\pi/2$ about the $Y$-axis to align with the basis definition (cf. Fig.~\ref{fig:rendering_basis}); this is applied just before the analysis step. By averaging, we merge two of the closest semi-axis lengths into one in order to find initial guesses for semi-axis lengths $a$ and $c$, based on which we classify whether the surface is oblate or prolate. 
\par
After the canonical transformation and spheroidal classification, we can now use a least-squares fit to find the spheroidal parameters $a$ and $c$ that best describe the surface. To this end, we write the general equations for oblate and prolate surfaces \textit{implicitly} as level-set functions:
\begin{equation}
    f_{oblate}(x, y, z) = c_{1} (x^2 + y^2) + c_{2} z^2 - 1.
\end{equation}
\begin{equation}
    f_{prolate}(x, y, z) = c_{1} x^2 + c_{2} (y^2 + z^2) - 1.
\end{equation}
We then find $c_1$ and $c_2$ from a least-squares fitting of the appropriate level-set function to the vertex matrix $P$. Afterward, the spheroidal coordinates can be extracted such that the focal distance $e = \sqrt{|1/c_1 - 1/c_2|}$, while $\zeta = \cosh^{-1}(a/e)$ or $\zeta = \sinh^{-1}(a/e)$ depending on whether the fitted spheroid is oblate or prolate, respectively. After reconstructing our surface with the spheroidal harmonics, we can reverse the rotations and placements removed from the canonical systems to match the registration of the input surface.

\subsubsection{Radial mapping onto spheroids for star-shaped (SS) particles}
\label{subsubsec:radial_mapping}
\par
The radial projection approach is straightforward to implement and computationally inexpensive, which explains its intensive use in particulate matter research. For a radial spherical parameterization onto $\mathbb{S}^2$, the radial mapping is obtained by simply normalizing the vectors pointing from the geometric centroid to any point on the surface. Such a normalization process maps all vertices on the surface to a unit sphere. For our parameterized radial map, we set the domain parameter $\text{AR}$ directly as the aspect ratio of the fitted canonical spheroid. We note that this combination of radial mapping and spheroidal analysis is equivalent to performing classical spherical harmonics on a \textit{rescaled} version of the granular particle, similar to the approach proposed in \cite{Huang2023} (but without their damping of the weights). %
\par
To ensure bijectivity, this radial-based mapping method requires that any radial vector pointing out from the geometric centroid intersects only once with the particle surface. In other words, the mapping should be always bijective, meaning that each point on the particle surface corresponds to a unique point on the target spheroid $\mathcal{E}$. If this is the case, the particle is characterized \say{star-shaped} (SS); otherwise, the particle is characterized \say{non-star-shaped} (NSS), similar to the use of these terms in SH approaches. Visual examples of SS and NSS particles are discussed in \ref{app:bijective_mapping}. In this work, we restrict the radial mapping approach described in this section to SS particles only.

\subsubsection{Hyperbolic mapping onto spheroids for star-shaped (SS) particles}
\label{subsubsec:nonlinear_radial_mapping}
\par
Following the exact spheroidal coordinate inversion, we can directly obtain $\eta$ associated with the hyperbolic curve intersecting a given location in $\mathbb{R}^3$. However, as explained above, the value of $\eta$ will depend on the choice of the domain parameter $e$. The examples in Fig.~\ref{fig:linear_vs_nonlinear_mapping_intro}A show that the choice of $e$ have a significant impact on the quality of the mapping. Specifically, when the focal points are outside of the convex hull of the surface points, the hyperbolic directors tend to cluster points in the vicinity of the major axes of the spheroid resulting in a trimmed spheroidal domain and often non-bijective images (see Fig.~\ref{fig:linear_vs_nonlinear_mapping_intro}A when $e = 2.5$). This results in a faulty reconstruction of the surface.
\par
\begin{figure}[!ht]
    \centering
    \includegraphics[width=0.45\linewidth]{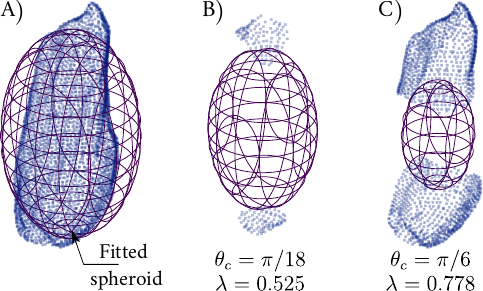}
    \caption{Comparison between the least-squares fit and the inscribed spheroids as a function of $\theta_c$ to find the target spheroid $\mathcal{E}$ for the stone FP\_C\_684\_1 [the stone was retrieved from \cite{dataset_1}]. (A) shows the results of the least squares fit with $\text{AR} = 0.4630$ (and the here redundant $e = 0.0805$) (prolate spheroid). (B) is the scaled spheroid results (using $\theta_c = \pi/18$) with an optimal $e_{opt} = 0.0626$. (C) same as (B) but with $\theta_c = \pi/6$ and a corresponding $e_{opt} = 0.0423$.
    }
    \label{fig:comp_lsq_opt_fit}
\end{figure}

To find a suitable choice of $e$ for a given particle, we start from the least-squares fit values of $a$ and $c$ and compute an initial guess for $e = \sqrt{|a^2 - c^2|}$. To find out if the value of $e$ extends beyond the particle surface, we define two cones emanating from the centroid to the North and South pole of the registered surface, respectively. Each cone has a user-defined opening angle $\theta_c$. Next, we filter out all the points that do not fall inside this cone, such that $|\theta| \notin (0, \theta_c) \text{ or} \notin  (0, \pi - \theta_c)$ for prolate surfaces and $|\pi/2 - \theta| \notin (0, \theta_c)$ for oblate ones. Here, $\theta = \tan^{-1} \left( z / \sqrt{x^2+y^2}\right)$ which is measured from the positive $Z$-axis. The last step is to iteratively shrink the fitted spheroid by shrinking $e$ to $\lambda e$ with $\lambda < 1$ until all the selected points by the cone fall outside the spheroid. Figure \ref{fig:comp_lsq_opt_fit} shows a comparison between the fitted target spheroid and the optimized ones as a function of $\theta_c$ and their corresponding scales $\lambda$. As can be seen from the same figure, the smaller $\theta_c$ the larger the target. In general, the reconstruction accuracy favors target spheroids that are as large as the input surfaces to reduce topological distortions. In this work, using $\theta_c = \pi/18$ was satisfactory for the tested oblate and prolate stones.
\par
After finding the optimal $e$, a particle may or may not be SS, as defined by whether any $\eta$ coordinate line intersects with the surface more than once. In \ref{app:bijective_mapping} we show examples of SS and NSS stones in the context of hyperbolic mappings. In the results below, we only apply hyperbolic mappings to SS surfaces.
\par
Overall, the reconstruction results of the herein-proposed optimal target spheroid with hyperbolic mapping were found to be better than using the proposed radial mapping with the fitted spheroid. This is mainly because the hyperbolic directors tend to closely sample points near the high curvature vicinity onto the target spheroids. This near-pole clustering helps in the regularization process and the convergence of the analysis. Further, the hyperbolic mapping tends to result in a more uniform sampling of $\eta$, which reduces the orthogonality error of the basis functions.  An example illustrating these differences between the two mapping approaches for a 2D problem is given in \ref{app:EFD_examples}.

\subsubsection{Conformalized mean curvature flow (cMCF) for fairing non-star-shaped (NSS) shapes}
\label{subsubsec:cMCF_mapping}
\par
Applying radial or hyperbolic mapping to NSS particles onto target spheroids results in nonbijective spheroidal images. To enable the ability to analyze NSS particles, we here propose a preprocessing step based on curvature flow iterations, in order to conformally map an NSS particle into an SS image, which enables the use of the radial or hyperbolic maps as proposed above. In general, the curvature flow methods resemble a smoothing (fairing) process that deforms the input surface into a more convex state.
\par
Several flow-based approaches exist for finding conformal mappings onto a target surface. The discrete Ricci flow \citep{Jin2008} approach is one of the most commonly used methods, with applications in, for instance, fairing rough surfaces. The Ricci flow however is an intrinsic flow method, so that after the flow reaches the target (user-defined) curvature we need to find the equivalent embedding into $\mathbb{R}^3$. A robust explicit flow alternative, more suitable to our context, is the conformalized mean curvature flow (cMCF) proposed by \cite{Kazhdan_2012}. This method is a modified version of the original MCF approach for convex surfaces \citep{Huisken1984}. A further improvement was proposed in \cite{Crane2013} to specifically enhance cMCF for surfaces with sharp features. However, the cMCF by \cite{Kazhdan_2012} was found sufficient for granular mechanics particles and will be briefly discussed below. We implemented the cMCF using the efficient finite element method (FEM) operators provided in \cite{libigl} that follow the discrete differential geometry operators defined in \cite{Meyer2003}.
\par
Unlike the Ricci flow approach, the cMCF does not flow into a prescribed target metric, instead, it converges into a sphere. Before converging into a sphere, the cMCF flow passes through intermediate diffeomorphism stages (fairing iterations) where the surface becomes star-shaped while the map remains conformal. Note that due to the conformal nature of the map, after many fairing iterations, we can obtain large local area distortions that can harm the surface reconstruction quality. In this work, we therefore generally perform iterations on NSS surfaces until they become star-shaped. After fairing we can apply one of the abovementioned radial or hyperbolic mappings as a complementary step to find the target spheroid of the diffeomorphism. Since the resulting diffeomorphism surfaces are close to being spheroidal in shape, applying the least-squares fitting approach of Section~\ref{subsec:fitting_spheroid} leads to a spheroid that is very close to the smoothed surface. In practice, the radial map performed similarly to the hyperbolic map on these smoothed particles. We therefore opt to forego the additional complexity of the hyperbolic mapping and combine the cMCF solely with the simple radial mapping approach in the results presented below.

\section{Results and discussions}
\label{sec:results}
\par
In this section, we present and discuss the results of the spheroidal harmonic approaches presented in this paper. We denote the proposed radial mapping in Section~\ref{subsubsec:radial_mapping} by rSOH and the hyperbolic one in Section \ref{subsubsec:nonlinear_radial_mapping} by hSOH, and the cMCF-preprocessed with radial mapping in Section~\ref{subsubsec:cMCF_mapping} by c-rSOH. We note that, as discussed above, rSOH and hSOH are only suited for SS particles, whereas the c-rSOH can be applied to both SS and NSS particles.
\par
First, we analyze common particles from the granular mechanics community where surfaces are usually convex or star-shaped (SS). For this, we compare the results of all three proposed approaches for two different datasets. We also compare our results with the traditional SH and show numerical evidence for the superiority of the proposed approaches. Second, we handle non-star-shaped (NSS) particles using the c-rSOH approach only. The datasets that we base our analysis on are summarized as follows: 
\begin{itemize}
    \item The first dataset consists of scanned stone masonry rubble. They were used to construct as-built digital twins for stone walls \citep{Saloustros2023}. The dataset was acquired by a portable laser scanner and can be accessed from \cite{dataset_1}.
    
    \item The second dataset is a collection of scanned aggregate particles \citep{Thilakarathna2021}. It was produced using a novel 3D light scanner. The dataset can be accessed from \cite{dataset_2}.
    
    \item The third dataset consists of scanned Calcite and Kieselkalk ballast stones that are used as railways ballast rubble \citep{Suhr_2020}. It can be accessed from \cite{dataset_3}.
\end{itemize}
\par
For reproducibility reasons, we base our benchmark surfaces on three datasets that are published online. Moreover, we use the code names of the analyzed particles as given by the original authors of each dataset. For consistency, prior to our analyses, we remeshed all the obtained raw data (point clouds or triangulated surfaces) via the screened Poisson reconstruction approach \citep{Kazhdan2013} implemented in the open-source library MeshLab \citep{meshlab_2008}. Moreover, to make triangular meshes Delaunay, we used the edge flipping strategy described in \citep{LAWSON_1977}.

\subsection{Analysis and reconstruction of star-shaped (SS) stones via c-rSOH, hSOH, and rSOH}
\label{sec:res_radial}
\par
We start by studying the SS particles that are included in the first two datasets \citep{dataset_1, dataset_2}. We analyzed and reconstructed a set of ten stones for each dataset using the radial, hyperbolic, and conformal-preprocessed spheroidal harmonics as well as a traditional spherical harmonics (SH) approach for comparison.
\subsubsection{Dataset 1: Stone masonry rubble (SS)}
\par
The provided point clouds in the first dataset \citep{Saloustros2023} are generally sparse and do not give enough details about the roughness and texture of the stones. Consequently, we used only $n_{max} = 20$ degrees for analyzing and reconstructing this dataset. Choosing higher expansion degrees for the particles in this dataset makes the analysis prone to diverge in the regularization step, especially when the mapping domain has large area distortion. As a reconstruction base, for all the proposed approaches, we used an icosahedron mesh subject to three refinement cycles (with $642$ vertices and $1280$ faces). The icosahedron spheres were rescaled to create a spheroidal base mesh for the reconstruction. This leads to slightly distorted triangles in the base mesh, but we here ignore this error that makes the points slightly nonuniform for the reconstruction basis.
\par
\begin{figure}[!hb]
    \centering \includegraphics[width=1.0 \linewidth]{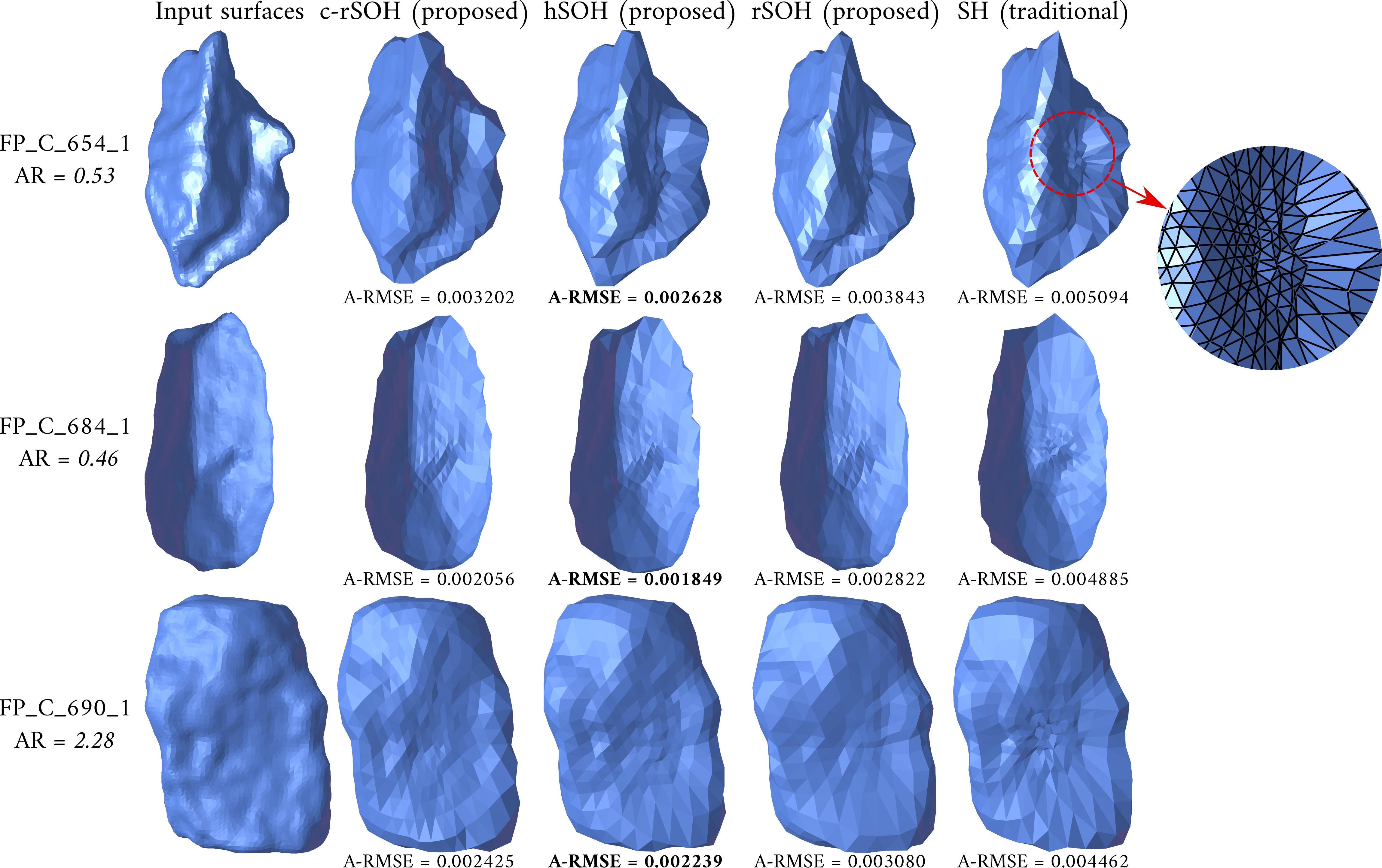}
    \caption{Reconstruction of a selected set of SS rubble stones \citep{dataset_1} via the conformal spheroidal approach (c-rSOH, second column), hyperbolic spheroidal approach (hSOH, third column), radial spheroidal approach (rSOH, fourth column), and traditional spherical harmonics (SH, fifth column). The code names and the fitted aspect ratios (AR) for the stones are shown on the far left annotations. Each reconstructed stone shows the normalized average Hausdorff distances (A-RMSE) below. Boldfaced \textbf{A-RMSE} highlights the best reconstruction accuracy.}
    \label{fig:dataset_1}
\end{figure}

Figure \ref{fig:dataset_1} shows a selected set of reconstruction results that serves as numerical comparisons among the c-rSOH, hSOH, rSOH, and SH approaches. The left column shows the remeshed input meshes from \cite{dataset_1}. The next four columns show the results of the (from the left) c-rSOH, hSOH, rSOH, and SH, respectively. For the c-rSOH approach, we only used three fairing iterations ($j=3$) and a non-dimensional time step $\delta = 0.0005$ with the cMCF method given that this time step is used for scaled diffeomorphisms with a surface area of unity. 
\par
We expressed the reconstruction error between the input and reconstructed meshes over a randomly sampled subset of points with the normalized average Hausdorff distances and denoted it by A-RMSE as annotated in Fig.~\ref{fig:dataset_1}. Here and in all results below we normalized A-RMSE with respect to the diagonal length of the bounding box (BB) of the input surface. In the same figure, boldfaced A-RMSE results highlight the best reconstruction results obtained across all the approaches. The rest of the results are shown in \ref{app:reconstruction_eg} and Fig.~\ref{fig:dataset_1_b}.
\par
The A-RMSE results show how the c-rSOH, hSOH, and rSOH outperform the traditional SH. The results obtained by the c-rSOH and hSOH approaches achieved the lowest A-RMSE values, with rather small differences between the proposed approaches for this dataset. The proposed approaches resulted in consistent triangular meshes that are not contracted and wavy in the middle band of the stones as those shown on the right side of the figures by SH. Notably, the oscillations resulting from SH dilute details in other parts of the stones that are far from the oscillations band (e.g., notice the reconstruction of the lower part of stone FP\_C\_785\_1 in Fig.~\ref{fig:dataset_1_b}).
\par
\begin{figure}[!hb]
    \centering
    \includegraphics[width=0.9\linewidth]{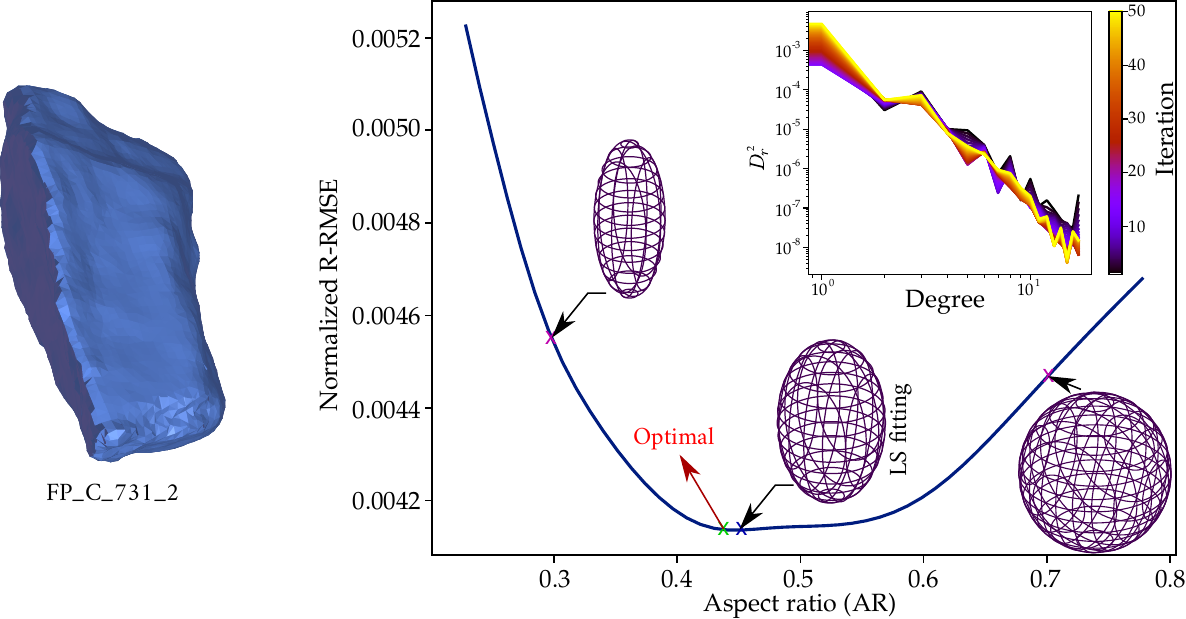}
    \caption{For stone $\text{FP\_C\_731\_2}$  \citep{dataset_1} (shown on the left) we vary the AR of the spheroidal domain, and record the normalized R-RMSE (plotted on the right). The optimal mapping was observed to be when $\text{AR} = 0.43$, while the one obtained from the least-squares fit proposed in Section~\ref{subsec:fitting_spheroid} was  $0.45$. The inset of the plot further shows the variation of the shape descriptors as we change the AR in 50 steps from $0.23 \ \to \ 0.78$}
    \label{fig:RMSE_vs_AR_ex}
\end{figure}

As discussed in Section \ref{subsubsec:radial_mapping}, the rSOH approach is AR-dependent. For each of the presented results of rSOH, we used the least-squares fit discussed in Section~\ref{subsec:fitting_spheroid} to set the aspect ratio used in the mapping. To study the impact of varying this domain parameter (AR) on the reconstruction accuracy, we manually varied the aspect ratio over a wide range of values and performed the rSOH method for each aspect ratio independently. Figure~\ref{fig:RMSE_vs_AR_ex} depicts the relation between the reconstruction radial root-mean-square error (R-RMSE) as a function of the chosen AR for a specific stone (see Fig.~\ref{fig:RMSE_vs_AR_mass_dataset_1} for the same analysis performed on other stones in the database). The R-RMSE is defined here as:
\begin{equation}\label{eqn:R_RMSE}
    \text{R-RMSE} = \frac{\sqrt{\sum_{i=1}^{n_v} \left(d^{(i)}_{OV, 1} - d^{(i)}_{OV, 2}\right)^2 / n_v}}{d_{diag}},
\end{equation}
where $d^{(i)}_{OV, 1}$ and $d^{(i)}_{OV, 2}$ are the radial distances from vertex $i$ to the centroid of each mesh, for the first and second meshes, and $d_{diag}$ is the diagonal length of the parallelepiped bounding box. The R-RMSE measure is used when the number of the vertices of both meshes matches and they are ranked in the same order. From Fig.~\ref{fig:RMSE_vs_AR_ex}, we can see that the optimal AR value does not exactly coincide with the least-squares fitted AR, though it is very close. This is also true for the other stones in the dataset, as shown in Fig.~\ref{fig:RMSE_vs_AR_mass_dataset_1}, indicating that our fitted AR achieves a near-optimal reconstruction for radially mapped SS particles. Further, note that in all these cases the R-RMSE significantly increases away from its minimum as the AR approaches unity, i.e.\ when approaching the SH results. Further, note that in all these cases the R-RMSE significantly increases away from its minimum as the AR approaches unity, i.e.\ when approaching the SH results, emphasizing the benefit of the SOH approach over the traditional SH method.
\par
The inset in Fig.~\ref{fig:RMSE_vs_AR_ex} plots the spectrum of shape descriptors for a range of spheroid aspect ratios. Here the radial shape descriptors $D_r^2$ are defined simply as the unweighted norm of the Fourier weights in the different coordinate directions:
\begin{equation} \label{eqn:shape_descriptors}
    D_r(n)^2 = D_x(n)^2 + D_y(n)^2 + D_z(n)^2, \qquad\text{with}\; {D}_{i}(n) = \sqrt{\sum_{m = -n}^{n} ||A^n_{m,i}||^{2}}, \quad \forall i \in \{x, y, z\}.\qquad
\end{equation}
\par
The plot demonstrates that the spectral properties of the reconstructed shape, as expressed in these shape descriptors, vary depending on the AR. This implies that these descriptors can not directly be used to compare stones parameterized with different ARs or compute fractal dimensions whenever we use the rSOH approach. Moreover, we see that the tail of the descriptors chart tends to stray away from zero for parameterization with ARs that are far from the optimal one; hence, showing a divergence behavior for the series.

\subsubsection{Dataset 2: Aggregate particles (SS)}
\par
\begin{figure}[!hb]
    \centering
    \includegraphics[width=1.0 \linewidth]{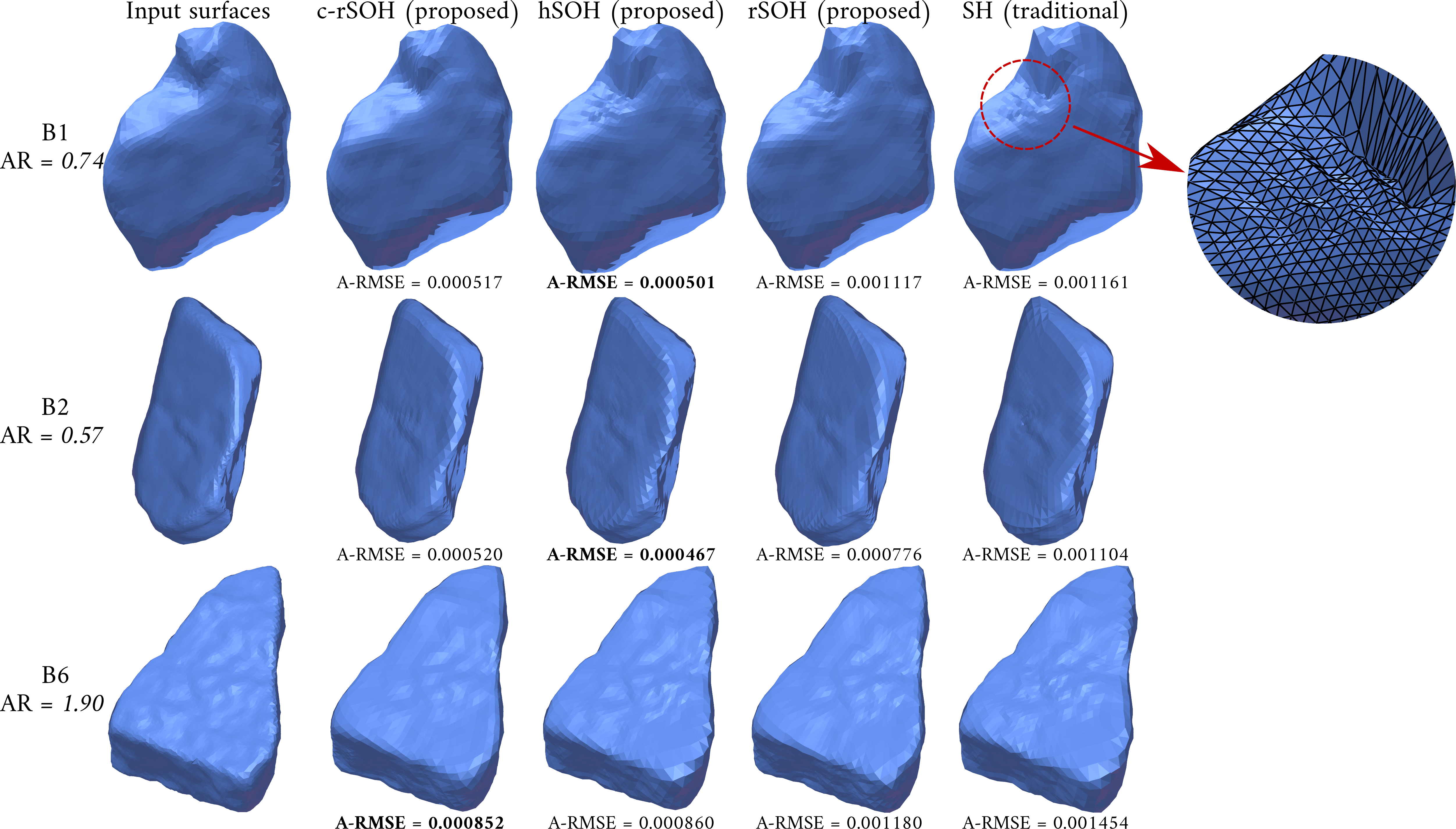}
    \caption{Reconstruction of a selected set of SS aggregates \citep{dataset_2} through the conformal- and radial-based approaches. The first column contains the input surfaces, while the second, third, and fourth are the reconstruction via the spheroidal harmonics c-rSOH, hSOH, and rSOH, respectively, as well as the traditional spherical harmonics (SH) in the fifth column. The left-hand texts show the annotation per stone and the fitted aspect ratio (AR) for the spheroidal coordinates. Each reconstructed stone shows the normalized average Hausdorff distances (A-RMSE) below.  Boldfaced \textbf{A-RMSE} highlights the best reconstruction accuracy.}
    \label{fig:dataset_2}
\end{figure}

The second dataset \citep{dataset_2} consists of a denser scan of aggregates. As this dataset provides higher frequency content, we used $n_{max} = 30$, apart from the B4 and B7 stones where we used $n_{max} = 20$. For the c-rSOH we used three fairing cycles for all the stones except for B7, which required five iterations to fair into a star-shaped surface. Figure~\ref{fig:dataset_2} shows the reconstruction results for all methods considered in this work, using four refinement cycles on an icosahedron as a reconstruction base (with $2562$ vertices and $5120$ faces). More results on this dataset are shown in Fig.~\ref{fig:RMSE_vs_AR_ex_dataset2} and \ref{fig:RMSE_vs_AR_mass_dataset_2} for the variation of the AR.
Similar to the first dataset, the results of the c-rSOH and the hSOH presented the best reconstruction accuracy in comparison with rSOH and the traditional SH. Although less than the SH, we see that a few stones constructed with rSOH and hSOH still show some minor oscillations (see Fig.~\ref{fig:dataset_2_b} for stones B4 and B7).
\par
For the second dataset, the hSOH significantly outperforms rSOH on the A-RMSE error metric. As the hSOH tends to sample denser near the poles of high curvature than the rSOH, we observed that the hSOH is, in general, more area-preserving than the rSOH.
\par
The problem of the large discrepancy between the surface point cloud and the target spheroid can be completely overcome using the c-rSOH approach. Overall, for star-shaped particles the hSOH and rSOH approaches provide improvements over SH with a little additional complexity. Out of the two, hSOH generally outperforms rSOH by reducing distortion near regions of high curvature. The c-rSOH is more general and can be used for SS stones as well, but requires additional computation cost while offering relatively little improvement over the other two methods. For our laptop computer with an Intel Core i7-11370H CPU, this additional cost is less than a second for stones consisting of $\mathcal{O}(10^4)$ vertices. However, this cost increases for relatively denser and more complex meshes.

\subsection{Analysis and reconstruction of non-star-shaped (NSS) stones via c-rSOH}
\label{subsec:NSS_stones}
\par

The third dataset consists of railway ballast stones, which pose the most challenges for the analysis and reconstruction tasks. The acquired meshes are denser, NSS, and can have sharp edges \citep{Suhr_2020} that require higher expansion degrees. The vicinity of sharp edges of the particles requires more points to describe abrupt changes in slope accurately. The harmonic reconstruction of the edges normally requires a higher reconstruction degree than the bulk of the surface. Such surfaces had reconstruction issues with the traditional SH approach and researchers tried to find alternative approaches to overcome these reconstruction challenges \citep{Ouhbi2017}.
\par
\begin{figure}[!hb]
    \centering
    \includegraphics[width=1.0 \linewidth]{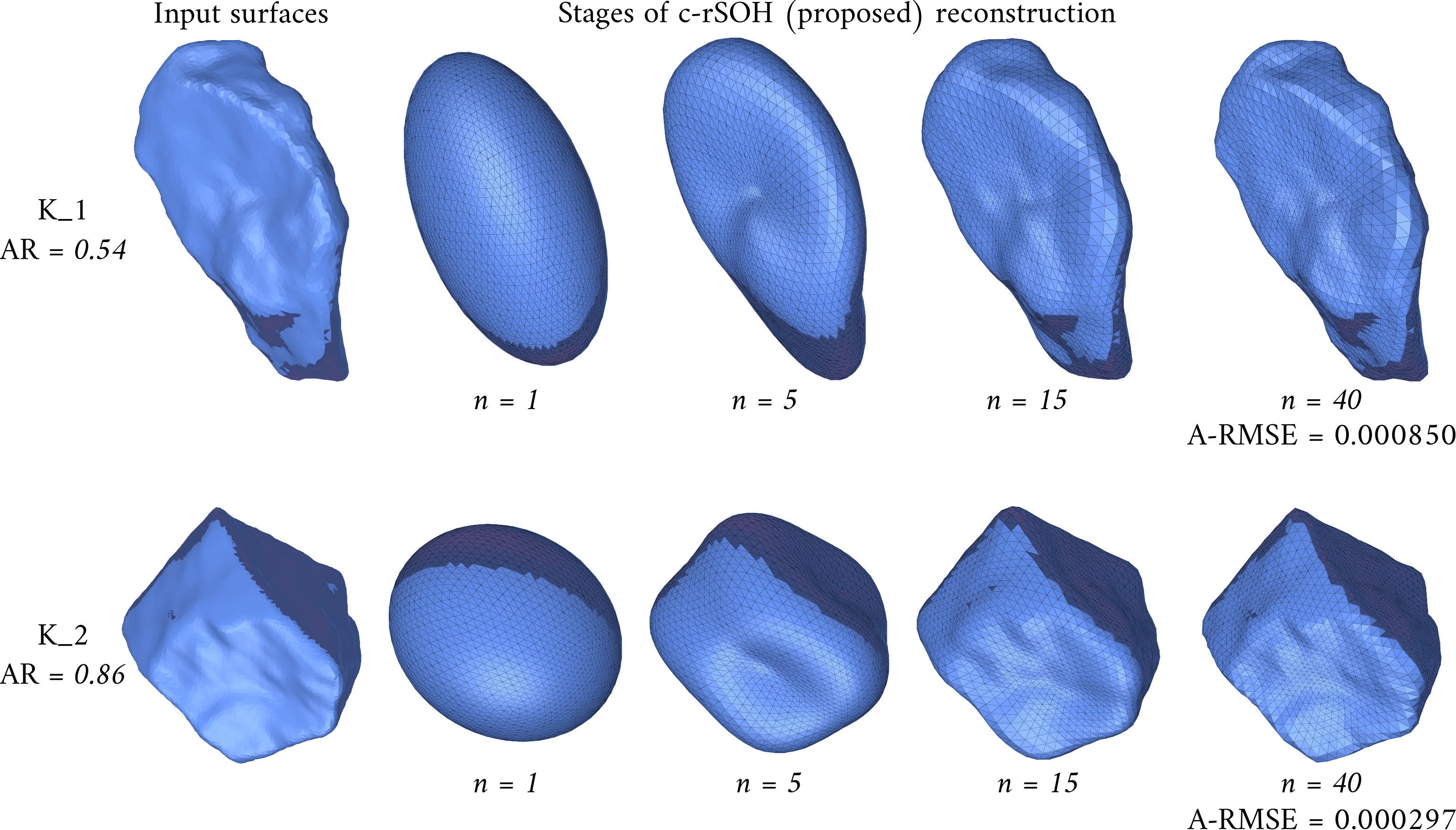}
    \caption{Reconstruction of a selected set of NSS stones \citep{Suhr_2020} through the c-rSOH approach. The conformal diffeomorphisms were obtained by setting the number of fairing iterations $j = 3$, and we show reconstruction stages up to $n_{max} = 40$ degrees. The final reconstructed stones are annotated (lower right) by the normalized average RMSE (A-RMSE). Intermediate reconstruction stages are also shown for $n \in \{1, 5, 15\}$. %
    }
    \label{fig:dataset_3}
\end{figure}

Using the c-rSOH approach, however, we successfully reconstructed all the surfaces with an accurate representation of the sharp edges at a relatively low computational cost. The successful reconstruction requires a combination of choosing the right reconstruction wavelength (related to $n_{max}$) and a proper reconstruction mesh size to avoid the Gibbs phenomenon. We conducted the c-rSOH analysis using $n_{max} = 40$ and three iterations with the cMCF mapping. For the reconstruction base, we used four refinement cycles of scaled icosahedrons. Figure \ref{fig:dataset_3} shows the reconstruction results of two stones as well as the actual input surface meshes. %
More reconstruction results for the c-rSOH approach are depicted in \ref{app:reconstruction_eg} for the rest of the dataset stones. The relation between the number of smoothing iterations and the reconstruction error is discussed in \ref{app:aspect_ratio_error}.

\subsection{Beyond granular mechanics applications}
\par
To demonstrate that the presented harmonic decomposition approaches can further be used beyond the intended scope of this paper, a complex visual benchmark was added in \ref{app:max_reconstruction}. Here we chose to analyze and reconstruct the surface of a Max Planck bust using the c-rSOH approach. In contrast, a c-SH approach (cMCF-preprocessed SH analysis) fails in this example. Consequently, we believe that our proposed SOH reconstruction method has applications beyond granular mechanics, for instance in medical or computer graphics fields \citep{Huang2005, Styner2006, Shen2009, Medyukhina2020, Grieb2022}.

\section{Conclusions} 
\label{sec:conclusions}
\par
To study the morphology of closed surfaces, we proposed the spheroidal harmonics (SOH) to generalize the traditional spherical harmonics (SH) approach. Our proposal expands the spherical parameterization domain $\mathbb{S}^2$ into a spheroidal one $\mathcal{E}$ that can be either oblate, prolate, or spherical with arbitrary dimensions for the two semi-major axes ($a$ and $c$). The newly added degrees of freedom allow for a more flexible parameterization domain that reduces the geometric distortions associated with mapping the particle and avoids the diverging analysis at high expansion degrees, thus, enhancing the analysis and reconstruction quality. As we use the associated Legendre polynomials and Fourier basis to construct the basis functions, the spheroidal harmonics (SOH) approach is similar to the spherical harmonics (SH). Such similarity will facilitate a seamless integration of this method within existing pipelines.
\par
We proposed three methods for parameterizing closed surfaces onto the spheroidal coordinates. The first method inverts the spheroidal coordinates to achieve a hyperbolic map (hSOH), whose domain is parameterized by the focal distance. The second method uses a rescaled radial-based projection similar to the SH approach (rSOH) so that the aspect ratio is the domain parameter. Between these two methods, the hSOH tends to sample points denser in the vicinity of regions of high curvatures, which has a favorable effect on the reconstruction results compared to rSOH. Moreover, the hSOH was observed to be area-preserving more than the rSOH. Both of these approaches are only valid for SS particles.
\par
Further, we proposed a third approach based on fairing input surfaces conformally, finding smooth diffeomorphism shapes of the input surfaces that can be subsequently mapped onto a spheroid via the radial mapping approach (c-rSOH). To perform fairing we used the conformalized curvature flow (cMCF) approach, which preserves the angles (conformal) of the input manifold. In contrast to rSOH and hSOH, this c-rSOH method works with both SS and NSS particles.
\par
We tested the rSOH, hSOH, and c-rSOH on three datasets and compared the results with the SH. The three proposed SOH-based approaches have successfully reconstructed all the SS particles from the tested datasets, while the traditional SH failed to reconstruct most of these particles without major oscillations. Overall, rSOH and hSOH provide low-cost alternatives to the traditional radial SH harmonics for SS particles, with the hyperbolic mapping method typically outperforming the radial mapping approach. The c-rSOH approach was used to analyze and reconstruct both SS and NSS particles at a small pre-processing computational cost; less than a second of additional computational time for stones of $\mathcal{O}(10^4)$ vertices on a standard laptop. For NSS particles, the c-rSOH is the only method that can perform the analysis; for SS particles the reconstruction quality of the c-rSOH approach is comparable to hSOH.
\par
The results of the proposed approaches give new avenues for improving the quality of morphology analysis of complex shapes. One possible extension of these methods is to combine angle-preserving with area-preserving approaches to obtain more uniformly spaced points on the target spheroidal manifold. Further, one can extend such approaches to deal with open surfaces to help quantify roughness and fractality on granular particles, as well as perform analysis in other physical and medical domains.

\section*{Reproducibility}
\par
To facilitate the reproducibility of the herein presented results, we made all the Python3.8 codes openly available under the GNU license on our repository:
\begin{itemize}
    \item GitHub: \url{https://github.com/msshaqfa/spheroidal-harmonics}
\end{itemize}
We refer readers to the cited datasets for permission reasons, despite having remeshed and cleaned the meshes for analysis.

\section*{Acknowledgement}
\par
The first author wants to thank the Swiss National Science Foundation (SNSF) for funding this work under project No. P500PT\_211088. The authors thank the members of the Earthquake Engineering and Structural Dynamics Laboratory (EESD) at EPFL especially Prof. Katrin Beyer, Dr. Savvas Saloustros, and Andrea Cabriada Ascencio for providing their dataset for the scanned stones. Thanks to Dr. Gary P.T. Choi (The Chinese University of Hong Kong CUHK) for the useful discussions and exchanged conversations.

\clearpage
\appendix

\section{Inverse coordinates and bijective mapping}
\subsection{Analytic inverse spheroidal coordinates}
\label{APP:inverse_coordinates}
\par
For completeness, this section shows the derivation of the exact oblate and prolate spheroidal inverse coordinates. This is important for computing the proposed hyperbolic mapping. Analogous to radial vectors in spherical coordinates, the normal directors of a spheroid are hyperbolic sections. For radial mappings, we can normalize the directors of each point on the input surface resulting in a spheroidal image of the surface denoted $\mathcal{E}$. In this paper, we use this analytic approach to normalize the directors and compute the parameterization of a surface onto a target spheroid.
\par
On the complex plane, we assume that the oblate coordinates are connected by the relation $\rho_{ob} + i z$, where $\rho_{ob} = \sqrt{x^2 + y^2}$. Now, we substitute the parametric coordinates in Eq.~\eqref{eqn:oblate_para1}. We can rewrite $\rho_{ob}$ for the oblate coordinates as $\rho_{ob} = e \ \cosh \zeta \cos\eta$ and it follows from that:
\begin{equation} \label{eqn:comple_plane_oblate}
    \rho_{ob} + i z = e \ \left( \cosh\zeta \cos\eta + i \sinh\zeta \sin\eta \right).
\end{equation}
We also know that the trigonometric and hyperbolic functions can be written in terms of Euler's formula such as:
\begin{align}
    \cos x = \frac{e^{ix} + e^{-ix}}{2},
    \sin x = \frac{e^{ix} - e^{-ix}}{2 i},\\
    \cosh x = \frac{e^{x} + e^{-x}}{2},
    \sinh x = \frac{e^{x} - e^{-x}}{2}.
\end{align}
By comparison, we can also write $\sinh x = -i \sin(i x)$ and $\cos x = \cosh (i x)$. Substituting all these identities in Eq.~\eqref{eqn:comple_plane_oblate} we obtain:
\begin{equation}
    \rho_{ob} + i z = e \cosh(\zeta + i \eta).
\end{equation}
The latter can be used for computing both $\zeta$ and $\eta$ coordinates such that:
\begin{equation}
    \label{eqn:analytic_inv_cood_oblate}
    \eta = \Im\left\{\cosh^{-1}\left( \frac{\rho_{ob} + i z}{e}\right)\right\},
    \quad 
    \zeta = \Re\left\{\cosh^{-1}\left( \frac{\rho_{ob} + i z}{e}\right)\right\}.
\end{equation}
\par
Similar to the above, we can derive the prolate coordinates. Though, for the prolate spheroids the complex plane is rotated by $\pi/2$; see Fig.~\ref{fig:oblate_vs_prolate}. So, the complex plane relation for prolate coordinates becomes $i \rho_{pr}(\eta, \phi) + z$. Similarly, we obtain:
\begin{equation}\label{eqn:analytic_inv_cood_prolate}
    \eta = \Im\left\{\cosh^{-1}\left( \frac{i \rho_{pr} + z}{e}\right)\right\},
    \quad
    \zeta = \Re\left\{\cosh^{-1}\left( \frac{i \rho_{pr} + z}{e}\right)\right\}.
\end{equation}
These analytic formulae can be used to compute the hyperbolic mapping onto any spheroid when normalizing the hyperbolic normal directors by $\zeta = \zeta_0$. This approach is also used as a starting point for the discussion in \ref{app:bijective_mapping}.

\subsection{Radial-based approximation of inverse spheroidal coordinates}
\label{app:linearized_inverse_coordinates}
\par
In this section, we derive a radial-based approximation of the inverse spheroidal coordinates given the analytic ones explained in \ref{APP:inverse_coordinates}. This radial inversion is used for computing $\hat{\eta}$-coordinates that we used for the radial mapping to replace the $\eta$-coordinates (hyperbolic curves). To demonstrate this, we here consider a simple, but not less general, case in the elliptic coordinates (Fig.~\ref{fig:oblate_vs_prolate}C) in the plane $\phi = 0$, which can be written in the parametric form:
\begin{align*}
    x(\zeta, \eta) &= e \cosh \zeta \cos \eta, \\
    z(\zeta, \eta) &= e \sinh \zeta \sin \eta,
\end{align*}
where $\zeta$ and $\eta$ are the exact spheroidal coordinates associated with Cartesian coordinates $x$ and $z$. In this radial mapping approach, we propose to normalize these coordinates such that:
\begin{align}
    \hat{x}(\zeta, \eta) &= \frac{x(\zeta, \eta)}{e \cosh \zeta_0} = \frac{\cosh \zeta}{\cosh \zeta_0} \cos \eta, \\
    \hat{z}(\zeta, \eta) &= \frac{z(\zeta, \eta)}{e \sinh \zeta_0} = \frac{\sinh \zeta}{\sinh \zeta_0} \sin \eta,
\end{align}
where $\zeta_0$ defines a \textit{target} elliptic section. From this, we can introduce the coordinate $\hat{\eta}$, such that:
\begin{align}
    \tan \hat{\eta} &= \frac{\hat{z}(\zeta, \eta)}{\hat{x}(\zeta, \eta)} \\
     &= \frac{\sinh \zeta}{\cosh \zeta}  \frac{\cosh \zeta_0}{\sinh \zeta_0} \frac{\sin \eta}{\cos \eta}\\
     &= \frac{\tanh \zeta} {\tanh \zeta_0} \tan \eta. \label{eqn:linearized_eta}
\end{align}
Rewriting $\zeta = \zeta_0 + \epsilon$ and expanding in $\epsilon$, we obtain :
\begin{align}
    \tan \hat{\eta} &= \frac{\tanh \left(\zeta_0 + \epsilon\right)} {\tanh \zeta_0} \tan \eta = \tan \eta \left[1 + \frac{\epsilon}{\sinh \zeta_0 \ \cosh \zeta_0} - \frac{\epsilon^2}{\cosh^2 \zeta_0} + \mathcal{O}(\epsilon^3) \right].
    \label{eqn:linearized_eta2}
\end{align}
\par
\begin{figure}[!ht]
    \centering
    \includegraphics[width=0.4\linewidth]{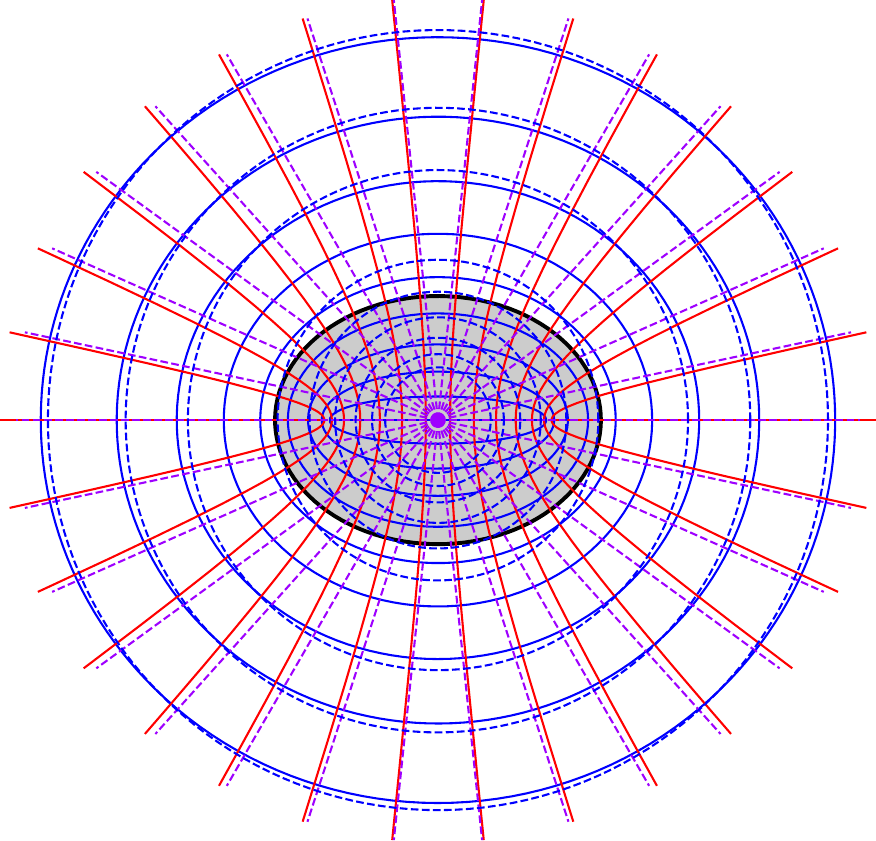}
    \caption{An overlay of the polar coordinates system on top of the elliptic one. The linear radials asymptotically approach the hyperbolic directors in the elliptic coordinate system. Solid lines represent the elliptic system with hyperbolic directors and confocal ellipses, while the dashed lines represent the radial directors with confocal circles.}
    \label{fig:radial_asympt_hyperbolic}
\end{figure}

From Eq.~\eqref{eqn:linearized_eta2}, we can see that for points that fall on the $\zeta_0$-ellipse (or spheroid in 3D) $\hat{\eta} = \eta$. For points away from the target shape the difference between $\eta$ and $\hat{\eta}$ increases with $\zeta - \zeta_0$. Consequently, choosing $\hat{\eta}$ instead of $\eta$ can be seen as a linearization (first-order approximation) in the parametric difference $\zeta - \zeta_0$. Figure \ref{fig:radial_asympt_hyperbolic} shows how the linear radials converge to the hyperbolic directors of the elliptic coordinates. In general, we approach polar coordinates from the elliptic ones when $e \to 0$ while $\cosh \zeta \to \sinh \zeta$.
The approach is generalized for spheroids by replacing ${x}(\zeta, \eta)$ by ${\rho}(\zeta, \eta, \phi)$ that can be either normalized by $e \cosh \zeta_0$ or $e \sinh \zeta_0$ for oblate and prolate surfaces, respectively, in the spheroidal space.

\subsection{On the star-shaped (SS) and non-star-shaped (NSS) particles and their bijectivity}
\label{app:bijective_mapping}
\par
\begin{figure}[!ht]
    \centering
    \includegraphics[width=.90\linewidth]{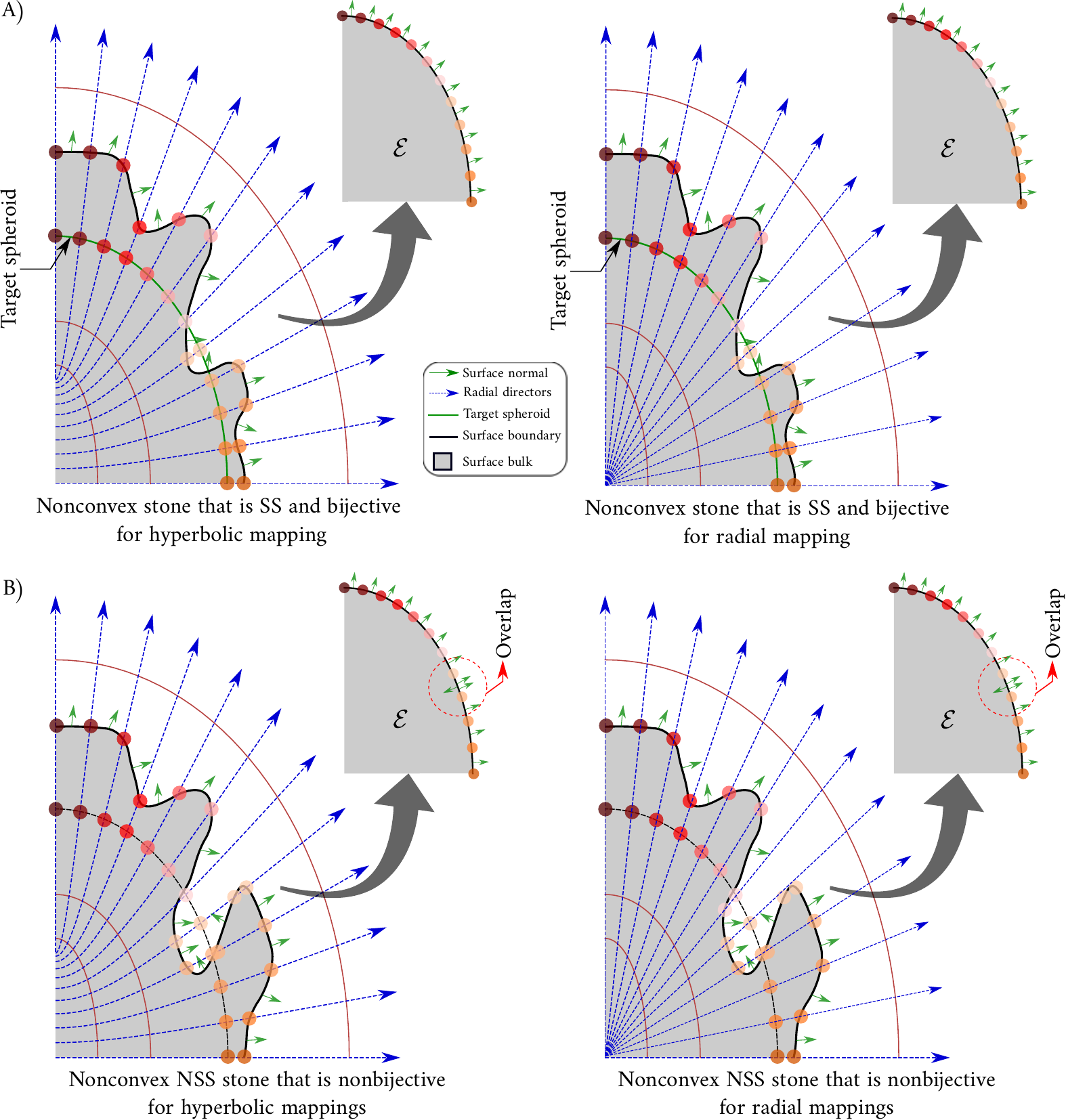}
    \caption{A cross-sectional view through two stones to illustrate the difference between star-shaped and non-star-shaped particles for the radial and hyperbolic mappings onto a spheroid. The color-coded dots on the \textit{target} spheroid are the result of sliding the sampled points from the discretized input surface along the normals of the spheroidal space (blue-dashed lines; hyperbolic sections for hyperbolic mapping and straight lines for the radial mapping). (A) and (B) explain the difference between the SS and NSS stones, respectively.
    }
\label{fig:star_shaped_test}
\end{figure}

For the radial and hyperbolic mapping methods, the distinction between star-shaped (SS) and non-star-shaped (NSS) surfaces is essential to ensure the bijectivity of the used mapping. Analyzing NSS surfaces may cause divergence in the spectral decomposition as the spheroidal image of the surface does not topologically match the original inputted surface. Figure \ref{fig:star_shaped_test}A and \ref{fig:star_shaped_test}B, show the difference between the definition of SS and NSS in the hyperbolic radial mapping (see \ref{APP:inverse_coordinates}) and the radial mapping (see \ref{app:linearized_inverse_coordinates}) methods through the use of an arbitrary but instructive \say{target spheroid}. The radial and hyperbolic mappings slide the surface sampled points onto the target spheroid along the corresponding normals shown with blue-dashed lines on the same figure. Assuming that the input surface normals (green arrows) always point outward of the stone bulk (away from the gray shading), then the spheroidal parameterization directors for SS surfaces will all point outward as well. If one or more of the normals flip direction, then the mapping of the triangulated surface is nonbijective or the surface is not manifold. This can also be expressed through the difference between the normal of the surface at a point, and the normal of the spheroid at the same point is larger than $\pi/2$, we have a nonbijective mapping and the surface is NNS. It should be noted that nonmanifold surfaces can cause normal flips, in this paper we assume that all the input meshes are manifold and the normals of the input surface are consistent (all point inwards or all point outwards). 
\par
To build on that, we use a simple numerical check for the bijectivity of discrete manifold surfaces. To numerically check that, we define a map $n: \mathcal{M} \to \mathbb{S}^2$, where $n$ is the Gaussian map and $n(p_i)$ is the normal vector at a given point $p_i$ that lies on $\mathcal{M}$. A surface is SS when all the normals are pointing outward (or all are inwards). The sign of the computed normals can be found via the signed area of each face. It should be noticed that for the hyperbolic mapping, the state of a given stone as SS or NSS is a function of the target spheroid (function of the scale $e$) that changes the direction of the normal hyperbolic directors.

\subsection{Radial and hyperbolic mappings with the elliptic Fourier descriptors (EFD)--2D counterpart examples}
\label{app:EFD_examples}
\par
To visualize the differences between the radial and hyperbolic mappings proposed here, we use the elliptic Fourier descriptor (EFD) approach \citep{Kuhl1982, Mollon2012} to study the reconstruction of 2D closed contours. Here we have generated an SS contour that resembles a 2D cross-section of a stone and sampled it uniformly with a set of points (constant arc length between the points).
\par
\begin{figure}[!hb]
    \centering\includegraphics[width=1.0\linewidth]{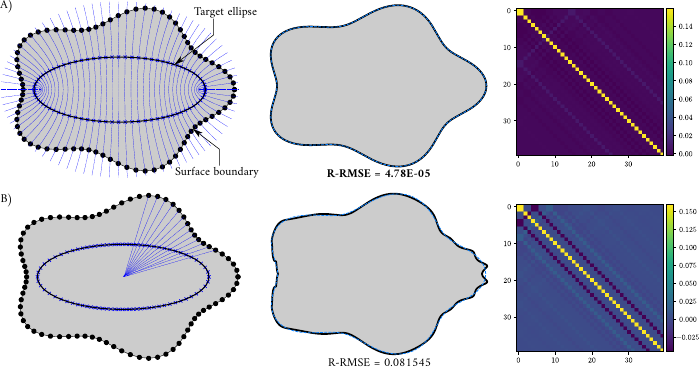}
    \caption{A comparison between the radial and hyperbolic mappings for 2D contours. (A) the left column shows the hyperbolic mapping onto an ellipse and the middle column is the corresponding reconstruction at $n_{max} = 20$ (the solid black line) as well as the right column is the orthogonality error matrix. (B) the left column same as the (A) but for the radial mapping. The blue-hashed lines in the middle column correspond to the original input surface contour. For this example, we used $141$ points (for the herein renders we used only $51$ points) to represent each contour and we sampled the $\eta$-coordinate with equal increments on each contour; the shown scatter on the left side does not represent the real count of the used points.}
    \label{fig:linear_vs_nonlinear_2D_contour}
\end{figure}

In Figure \ref{fig:linear_vs_nonlinear_2D_contour}, we show the results of mapping the points using a hyperbolic map (A) and a radial map (B). The mapped points are again visualized using an arbitrary \say{target ellipse}. Further, we use a similar least squares approach as in 3D to integrate the Fourier basis functions. This also allows us to further investigate the orthogonality of the discrete basis as a function of the sampling approaches.
The results, on the same figure, show that the hyperbolic mapping favors sampling points near the poles (high-curvature areas), while the radial mapping favors sampling denser far from the poles (low-curvature areas). The effect of these mappings is shown in the middle column of sets (A) and (B) where we see oscillations occurring near the pole areas in the radial mapping due to the poor regularization in that area and weak sampling. Moreover, the right column of the figure shows the corresponding orthogonality matrix for the constructed Fourier basis using $\eta$ for the hyperbolic mapping and $\hat{\eta}$ for the radial one where the error for the radial basis shows more off-orthogonality error.
\par
\begin{figure}[!hb]
    \centering
    \includegraphics[width=1.0\linewidth]{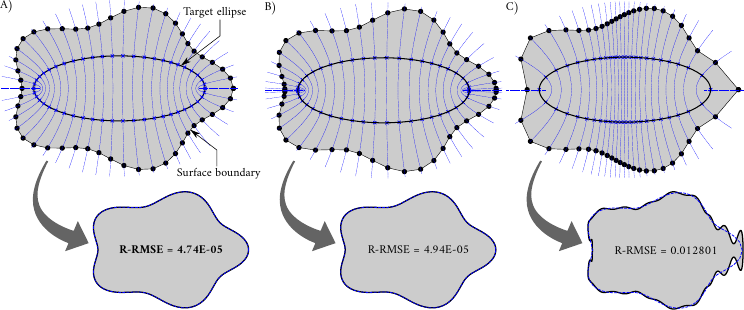}
    \caption{Comparing the effect of clustering sampled points in different regions. (A) uniformly sampled target spheroid with equal elliptic-arc length. (B) Densely sampling the target ellipse near high curvature areas (the poles). (C) Densely sampling the target ellipse near low curvature areas. The underneath insets show the reconstruction results using $n_{max} = 20$ and with $161$ sampled points for each contour. The herein-shown contours do not reflect the real number of sampled points and are mere visualization contours for the effect of sampling target ellipses.}
    \label{fig:sampling_2D_contour}
\end{figure}

To further investigate the effect of the point distribution after mapping, we artificially distributed the \textit{source} points in three different ways. In the first approach, we created source points so that the $\eta$ coordinate is uniformly sampled after mapping (see Fig.~\ref{fig:sampling_2D_contour}A). In the second approach, we created source points more densely near the pole area (see Fig.~\ref{fig:sampling_2D_contour}B). For the third one, we created most of the points near the low curvature areas (see Fig.~\ref{fig:sampling_2D_contour}C). After reconstructing the three cases, we find that the best results were obtained by the first approach. However, the results of the second approach were very close to the first one, while the third one deteriorated much faster for the same number of sampled points. This demonstrates that approaches that favors sampling more uniformly or even denser near the poles are in general better than those oversampling the low-curvature regions. This supports the main results in the manuscript that hyperbolic-based mapping is more beneficial than radial-based mapping. 
\par
\begin{figure}[!hb]
    \centering
    \includegraphics[width=0.9\linewidth]{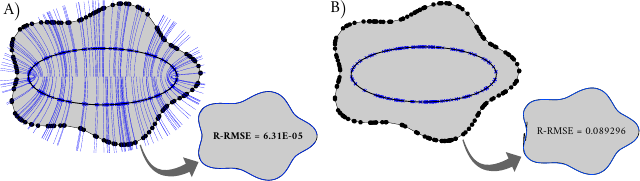}
    \caption{Reconstructing contours with radial and hyperbolic mappings of a sampled $161$ random points from the 2D contour of the surface with a fixed seed number $5$. (A) shows the randomly sampled stone using hyperbolic mapping and the resulting reconstruction. (B) same as (A) but with radial mapping.}
    \label{fig:random_contour_rec}
\end{figure}

Finally, to consider a more realistic example, we used random sampling on the input surface itself. By fixing the random seed number, we tested both the hyperbolic and radial sampling approaches onto the same target spheroid. Figure \ref{fig:random_contour_rec} shows the numerical differences between the hyperbolic and radial mapping and the arc-length distortion for the resulting target spheroid. The hyperbolic sampling was generally more preserving in terms of the arc length (equivalent to area-preserving in the 3D case). 

\section{Supplemental results}
\label{app:suppl_mat}

\subsection{Stones reconstruction examples}
\label{app:reconstruction_eg}
\par
\begin{figure}[!hb]
    \centering
    \includegraphics[width=0.90 \linewidth]{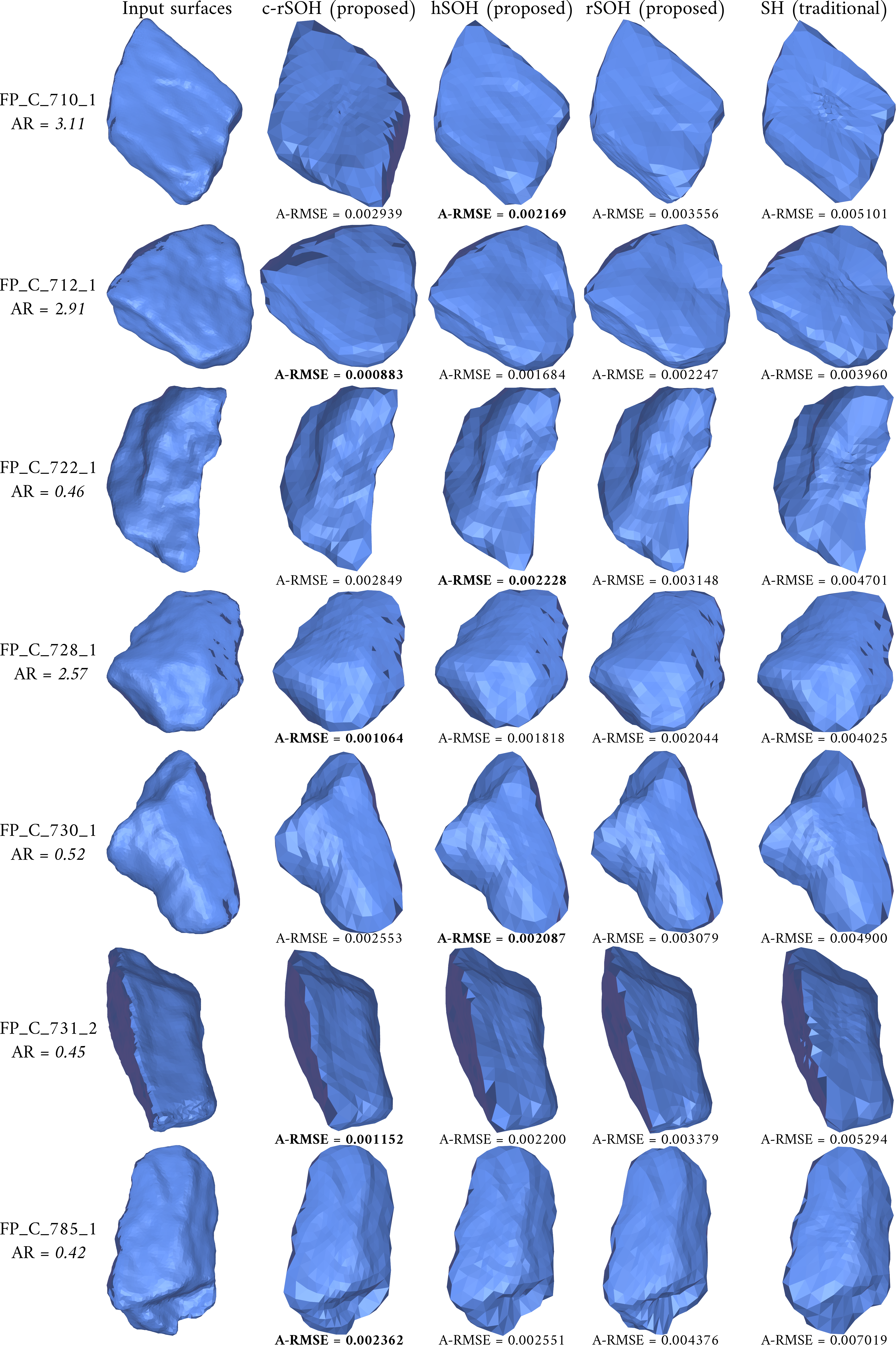}
    \caption{Reconstruction of a selected set of SS rubble stones \citep{dataset_1} via the conformal spheroidal approach (c-rSOH, second column), hyperbolic spheroidal approach (hSOH, third column), radial spheroidal approach (rSOH, fourth column), and traditional spherical harmonics (SH, fifth column). The codenames and the fitted aspect ratios (AR) for the stones are shown on the far left annotations. Each reconstructed stone shows the normalized average Hausdorff distances (A-RMSE) below. Boldfaced \textbf{A-RMSE} highlights the best reconstruction accuracy.}
    \label{fig:dataset_1_b}
\end{figure}
\begin{figure}[!hb]
    \centering
    \includegraphics[width=0.90 \linewidth]{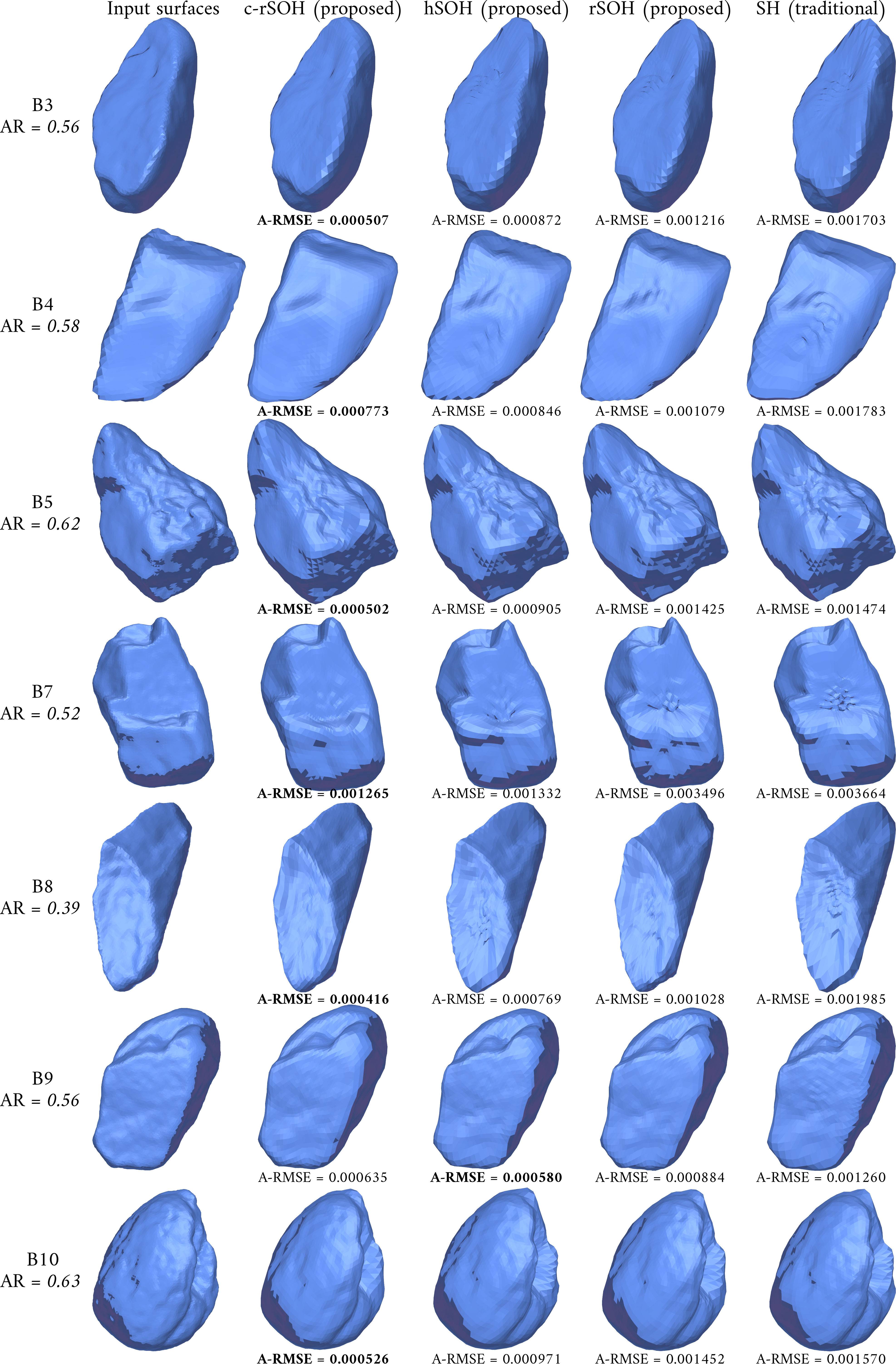}
    \caption{Reconstruction of a selected set of SS aggregates \citep{dataset_2} through the conformal- and hyperbolic as well as radial approaches. The first column contains the input surfaces, while the second, third, and fourth are the reconstruction via the spheroidal harmonics c-rSOH, hSOH, and rSOH, respectively, as well as the traditional spherical harmonics (SH) in the fifth column. The left-hand texts show the annotation per stone and the fitted aspect ratio (AR) for the spheroidal coordinates. Each reconstructed stone shows the normalized average Hausdorff distances (A-RMSE) below.  Boldfaced \textbf{A-RMSE} highlights the best reconstruction accuracy.}
    \label{fig:dataset_2_b}
\end{figure}

In this section, we present the rest of the results for the chosen stones from the three datasets. Figure \ref{fig:dataset_1_b} is a continuation of Fig.~\ref{fig:dataset_1} where we compare the reconstruction via the traditional SH and the newly proposed SOH-based approaches. Similarly, Fig.~\ref{fig:dataset_2_b} is also a continuation of Fig.~\ref{fig:dataset_2} to compare the reconstruction results between the same abovementioned approaches.
\par
\par
\begin{figure}[!hb]
    \centering
    \includegraphics[width=0.75 \linewidth]{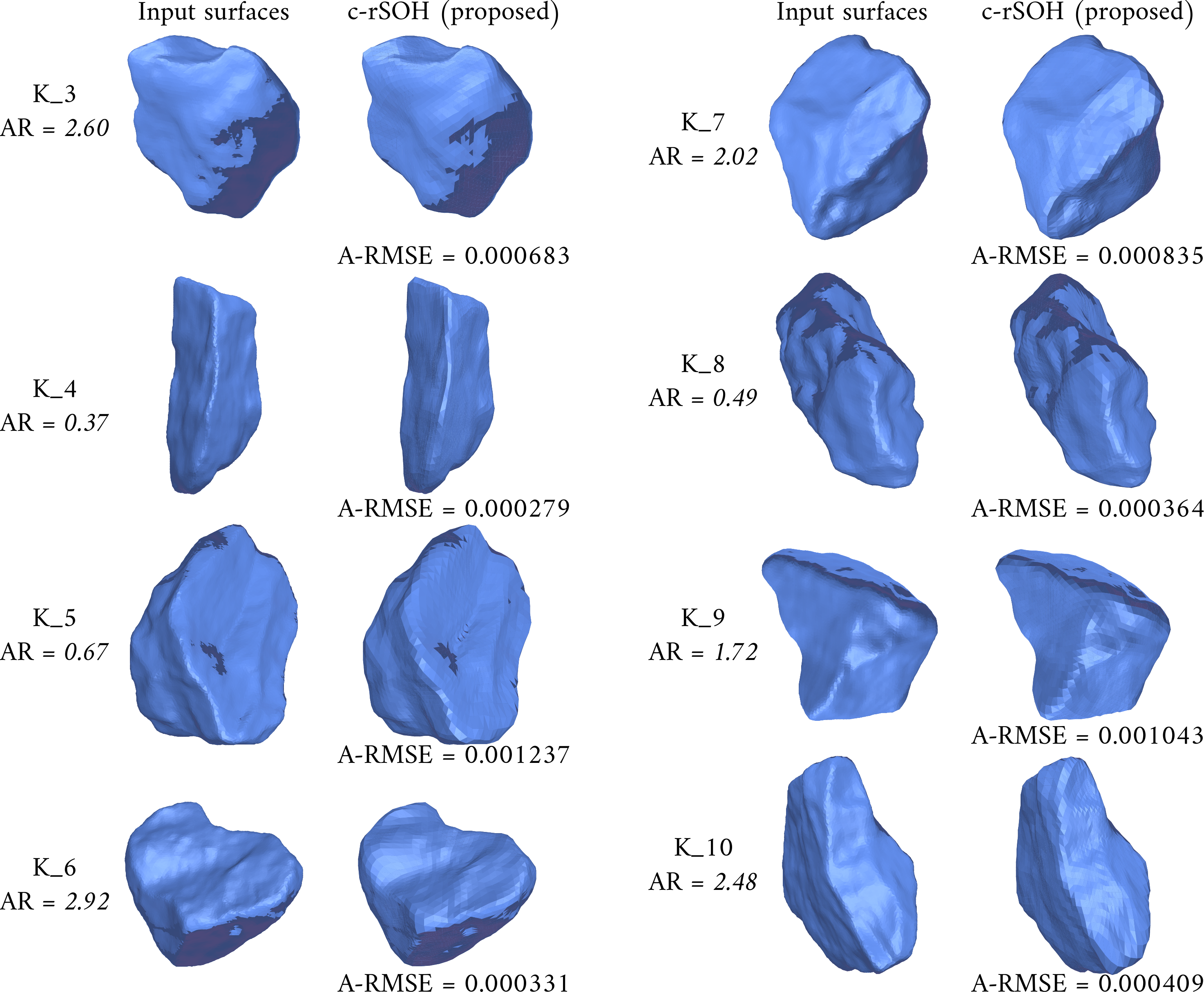}
    \caption{Reconstruction of a selected set of rubble stones \citep{Suhr_2020}. This figure is a continuation of Fig.~\ref{fig:dataset_3}, where the NSS railway ballast rubbles were reconstructed by the c-rSOH approach. Each reconstructed stone shows the normalized average Hausdorff distances (A-RMSE) below.}
    \label{fig:dataset_3_b}
\end{figure}

Figure \ref{fig:dataset_3_b} is a continuation of Fig.~\ref{fig:dataset_3} where we show the results of the reconstruction using c-rSOH approach. 

\subsection{Aspect ratio \texorpdfstring{$a/c$}{ } versus RMSE}
\label{app:aspect_ratio_error}
\par
\begin{figure}[!hb]
    \centering
    \includegraphics[width = 0.90\linewidth]{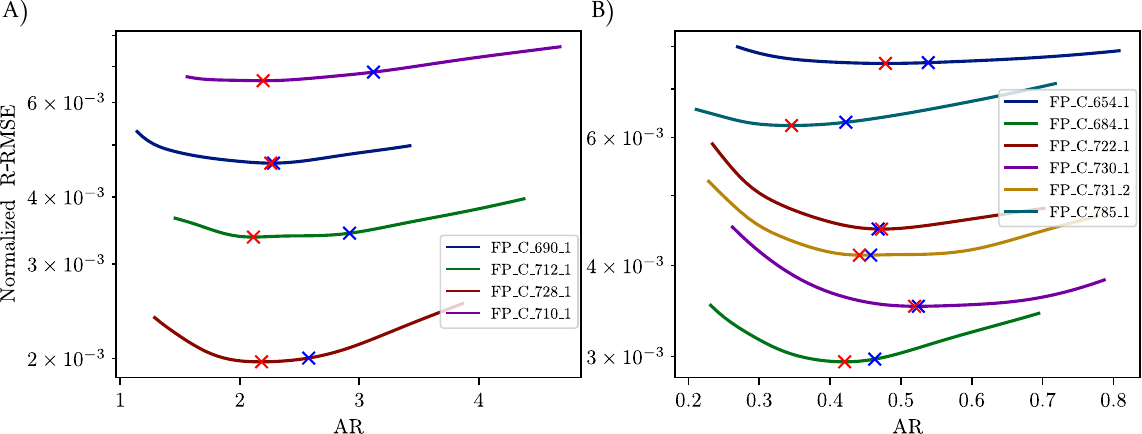}
    \caption{Optimization curves for R-RMSE vs. AR for the first dataset retrieved from \cite{dataset_1} and reconstructed via the rSOH in Fig.~\ref{fig:dataset_1} and ~\ref{fig:dataset_1_b}. (A) is the reconstruction for oblate-like particles $a/c > 1$ and (B) for prolate-like ones where $a/c < 1$. Red crosses mark the optimal AR and blue crosses are for the LS fit.}
    \label{fig:RMSE_vs_AR_mass_dataset_1}
\end{figure}

Figure~\ref{fig:RMSE_vs_AR_mass_dataset_1} shows the results of the reconstruction accuracy using the rSOH approach for stones of the first dataset (oblate (A) and prolate (B) particles) when varying AR (rSOH is AR-dependent). The estimates of the least-squares strategy, see Section \ref{subsec:fitting_spheroid}, give reasonable solutions for SS particles. From the figure, we notice that the closer the spheroid is to a sphere the larger the reconstruction error. Although not optimal, the spheroidal fitting provides a conveniently fast and good reconstruction making it a valid approach for practical usage for SS particles.
\par
\begin{figure}[!hb]
    \centering
    \includegraphics[width=0.9\linewidth]{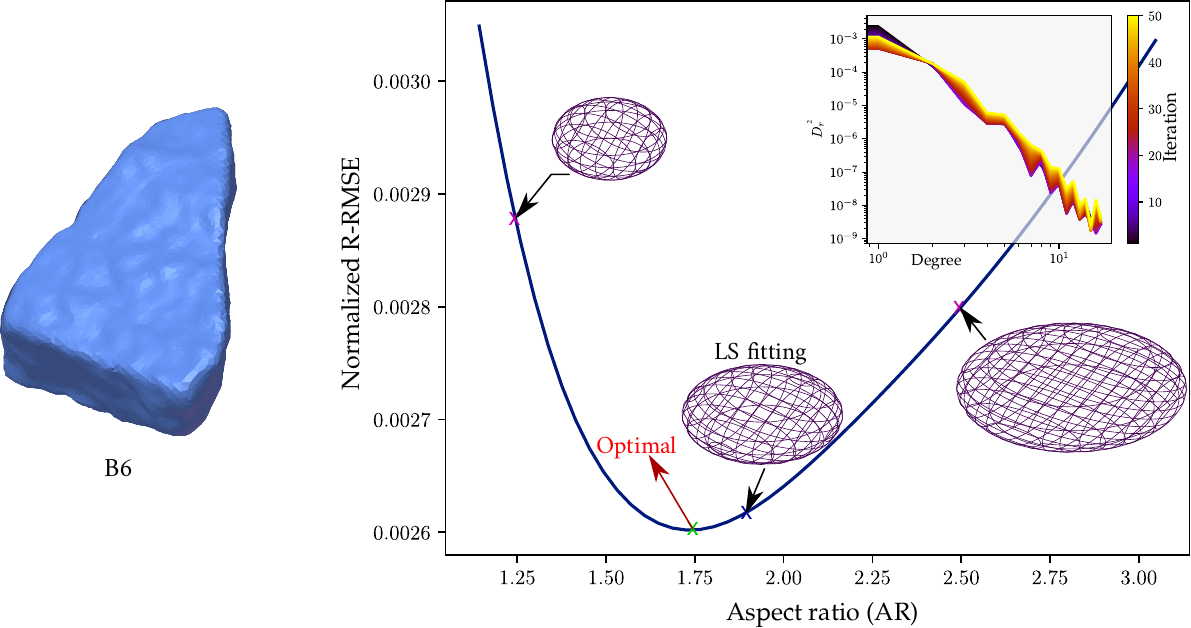}
    \caption{The radial normalized R-RMSE vs. the aspect ratio (AR) for stone B6 \citep{dataset_2} reconstructed via rSOH. The inset shows the variation of the shape descriptors as we change the AR from $1.14 \ \to \ 3.04$ (the interval was spaned by $50$ iterations). The optimal AR was observed to be $1.74$, while the one obtained from the least-squares fit was $1.90$. As we approach a sphere AR $= 1.0$ or AR $= 3.0$ the solution diverges faster. Notice that the optimal error is not $0.0$ due to the truncation error ($n_{max} = 20$) and the projection error for finding Fourier weights.}
    \label{fig:RMSE_vs_AR_ex_dataset2}
\end{figure}
\begin{figure}[!hb]
    \centering
    \includegraphics[width = 0.5\linewidth]{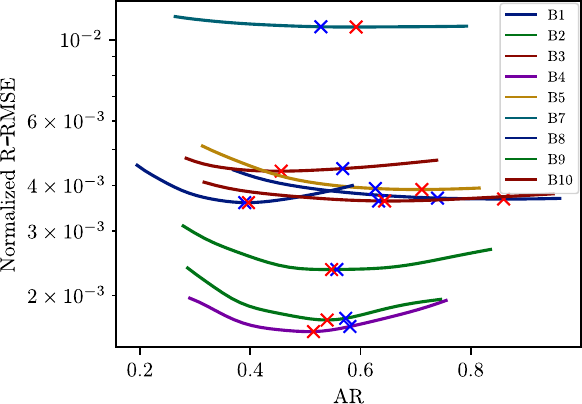}
    \caption{rSOH optimization curves for R-RMSE versus AR for the dataset retrieved from \cite{dataset_2} and reconstructed in Fig.~\ref{fig:dataset_2} and ~\ref{fig:dataset_2_b} for prolate-like particles where $a/c < 1$. Red crosses mark the optimal AR and blue crosses indicate the aspect-ratio of the least-squares fit.}
    \label{fig:RMSE_vs_AR_mass_dataset_2}
\end{figure}

Similar to the presented results of the first dataset, Fig.~\ref{fig:RMSE_vs_AR_ex_dataset2} shows an example of the R-RMSE variation with the rSOH approach along with the AR. This figure also explains how insignificant changes in the R-RMSE do reflect some considerable changes in the shape descriptors (see the inset of the same figure) that will undoubtedly affect the estimated fractal dimension of that stone.  
Figure~\ref{fig:RMSE_vs_AR_mass_dataset_2} summarizes the results for all the tested stones in the second dataset \citep{dataset_2}. The results, however, show some discrepancies between the fit and the optimal AR for the stones B3 and B7. This indeed corresponded to minor osculations as depicted in Fig.~\ref{fig:dataset_2_b} for the rSOH approach. Nevertheless, the reconstruction quality is still better than the traditional spherical harmonics with radial parameterization.
\par
\begin{figure}[!hb]
    \centering
    \includegraphics[width = 0.80\linewidth]{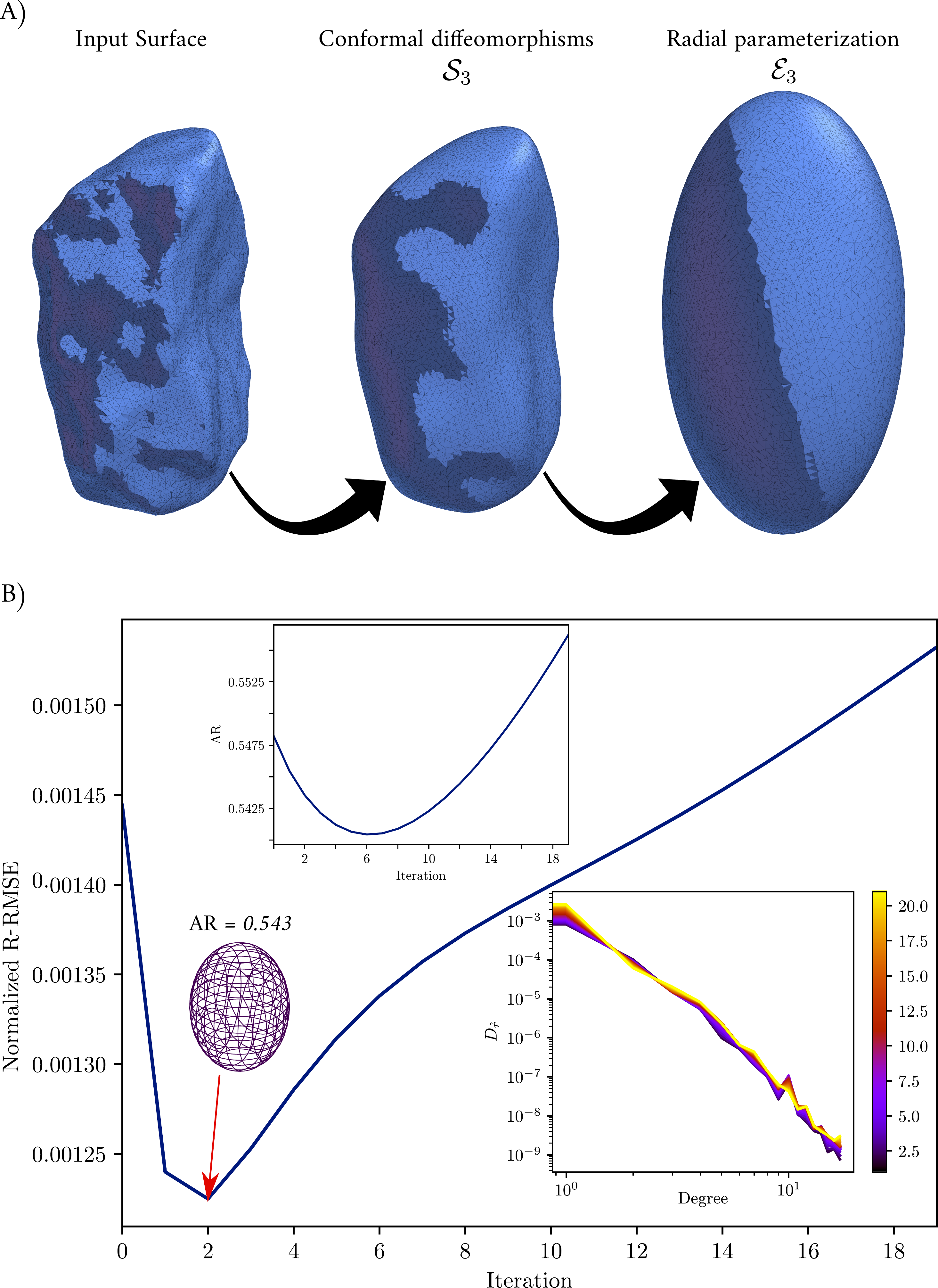}
    \caption{(A) depicts the conformal-based parameterization approach for the mesh of stone K\_1 \citep{dataset_3}; the same stone but different view from Fig.~\ref{fig:dataset_3}. (B) Optimization curves for R-RMSE vs. the curvature flow iteration with a time step $\delta =  0.0005$ for the dataset retrieved from \cite{Suhr_2020} and reconstructed in Fig.~\ref{fig:dataset_3} and \ref{fig:dataset_3_b}. The analysis results of this figure all belong to stone K\_1 from the third dataset.}
    \label{fig:RMSE_vs_itrs_mass_dataset_3_B1}
\end{figure}

For the third dataset, the error sensitivity was approached slightly differently from the former two datasets. We studied the effect of the maximum number of smoothing iterations used for the cMCF approach on the final reconstruction RMSE. Figure \ref{fig:RMSE_vs_itrs_mass_dataset_3_B1}(A) shows the stone K\_1 \citep{dataset_3} and the stages of mapping the stone (see Section \ref{subsubsec:cMCF_mapping}). Figure \ref{fig:RMSE_vs_itrs_mass_dataset_3_B1}(B) shows the relation between the cMCF iterations and the reconstruction accuracy of the stone. We can also see that the shape descriptors scatter is less than the former two datasets as the AR itself is not changing much due to iterations as shown in the two insets of the same figure.  
\par
\begin{figure}[!hb]
    \centering
    \includegraphics[width = 0.9\linewidth]{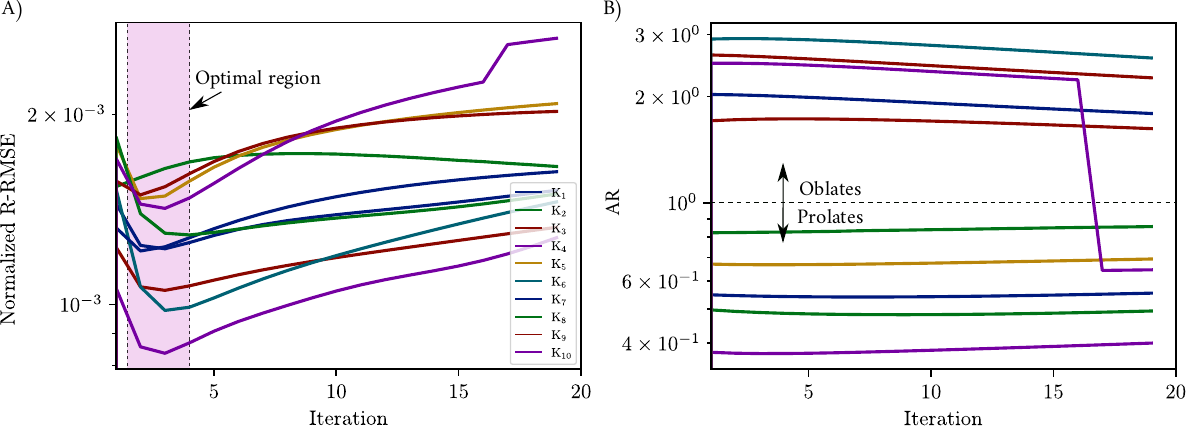}
    \caption{Optimization curves for the reconstruction error with respect to the maximum iterations used in the cMCF approach. (A) shows the radial normalized R-RMSE with the cMCF used iterations before applying the radial mapping and analysis process. The used time step here is $\delta = 0.0005$ for the dataset retrieved from \cite{Suhr_2020}. (B) shows the change of AR with iterations for oblate and prolate stones. The visual reconstruction results are depicted in Fig.~\ref{fig:dataset_3} and \ref{fig:dataset_3_b}.}
    \label{fig:RMSE_vs_itrs_mass_dataset_3}
\end{figure}

As shown in Fig.~\ref{fig:RMSE_vs_itrs_mass_dataset_3}(A), most of the stones show better results once we flow the surface with $2-4$ iterations only before applying the radial mapping. As a result, we obtained very good results for more challenging stones like the railway ballast rubbles. Fig.~\ref{fig:RMSE_vs_itrs_mass_dataset_3}(B) shows how the AR changes with iterations for both oblates (AR $>1$) and prolates (AR $< 1$).

\subsection{Conformal parameterization and reconstruction of a visual benchmark}
\label{app:max_reconstruction}
\par
\begin{figure}[!ht]
    \centering
    \includegraphics[width=1.0\linewidth]{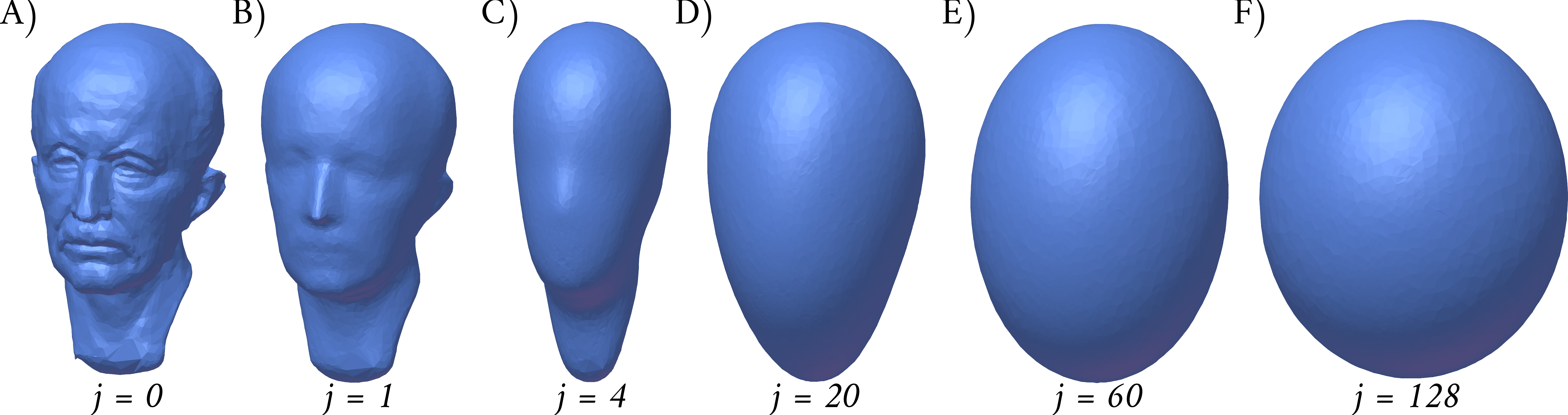}
    \caption{The flow of Max Planck bust via the cMCF parameterization \citep{Kazhdan_2012} at different $j$ iterations up to $128$ [insets (A) $\to$ (F)]. As the flow progresses, the surface converges into a sphere $\mathbb{S}^2$. The surfaces of each subset were rescaled to conserve the original surface area of the surface. Notice, at $j = 4$ [inset (C)] the surface became star-shaped and can be radially mapped into a spheroid (i.e., $\mathcal{E}_4$).}
    \label{fig:max_head_para}
\end{figure}

Figure \ref{fig:max_head_para} depicts the parameterization of the Max Planck bust benchmark surface. For this NSS surface, we used the cMCF approach \citep{Kazhdan_2012}. The step size (each time step is an iteration $j$) for the solver was set to $\delta = 0.0005$.
\par
As can be seen in the figure, the surface evolution passes through intermediate stages at each time step (diffeomorphisms) that are equivalent to the input surface (conformal equivalence). The further we flow the surface (more iterations) it converges into a sphere, however, with large area distortion. Over time, these almost-spheroidal intermediate shapes become convex and can be used as a point to radially map, see Section \ref{subsubsec:radial_mapping}, the surface into a spheroid without being NSS.
\par
\begin{figure}[!hb]
    \centering
    \includegraphics[width=1.0 \linewidth]{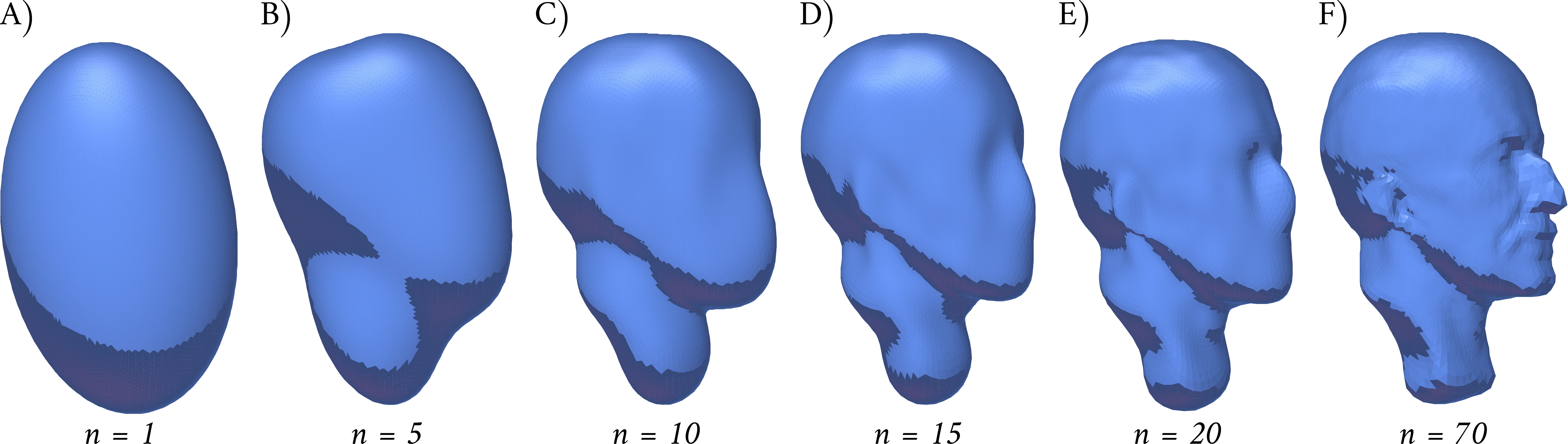}
    \caption{Reconstruction of Max Planck bust as a visual benchmark via the spheroidal harmonics. For the mapping, we used the cMCF whence $j = 10$ and $\delta = 0.0005$. With these settings, we were able to be convergent up to $n_{max} = 70$ degrees.}
    \label{fig:max_head_rec}
\end{figure}

Combining the cMCF approach with the radial projection mapping, see Section \ref{subsubsec:cMCF_mapping}, becomes handy for mapping NSS objects. Figure~\ref{fig:max_head_rec} shows the results of the harmonic decomposition of the herein-proposed visual benchmark. Similar to the SPHARM \citep{Brechbuhler_1995} and \cite{shaqfa2021b} the reconstruction at $n = 1$ results in an ellipsoidal shape that can be used for estimating the reconstruction wavelength.
\par
In Fig.~\ref{fig:max_head_rec}, we have flown the surface using the first ten time steps ($j = 10$), and after that, we applied the proposed radial mapping. We have determined this number of iterations as we obtain an optimal reconstruction at this number of iterations. Again, the process of finding an optimal spheroidal size is not trivial, and it is hard to find without posing the problem as an optimization problem. Finally, we include two videos in the supplemental material that show the cMCF flow and the reconstruction of the same surface.

\clearpage

\clearpage

\bibliographystyle{unsrtnat}

\bibliography{main}

\end{document}